\documentclass[pra,superscriptaddress,showpacs,amsmath,amssymb,twocolumn]{revtex4}
\usepackage{bm}
\usepackage{graphicx,color,ulem}
\usepackage{verbatim}

\begin{document}

\newcommand{\tr}{\mathop{\mathrm{tr}}\nolimits}
\newcommand{\adj}{\mathop{\mathrm{adj}}}
\renewcommand{\Re}{\mathop{\mathrm{Re}}}
\renewcommand{\Im}{\mathop{\mathrm{Im}}} 
\newcommand{\HA}{\mathop{\mathcal{H}}\nolimits} 
\newcommand{\D}{\mathop{\mathcal{D}}\nolimits}
\newcommand{\T}{\mathop{\mathcal{T}}\nolimits} 
\newcommand{\LA}{\mathop{\mathcal{L}}\nolimits} 
\newcommand{\B}{\mathop{\mathcal{B}}\nolimits} 
\newcommand{\C}{\mathop{\mathcal{C}}\nolimits} 
\newcommand{\SA}{\mathop{\mathcal{S}}\nolimits}
\newcommand{\E}{\mathop{\mathcal{E}}\nolimits}
\newcommand{\F}{\mathop{\mathcal{F}}\nolimits}
\newcommand{\U}{\mathop{\mathcal{U}}\nolimits} 
\newcommand{\V}{\mathop{\mathcal{V}}\nolimits} 
\newcommand{\M}{\mathop{\mathcal{M}}\nolimits} 
\newcommand{\R}{\mathop{\mathbb{R}}\nolimits} 
\newcommand{\CA}{\mathop{\mathbb{C}}\nolimits} 
\newcommand{\N}{\mathop{\mathbb{N}}\nolimits}
\newcommand{\I}{\mathop{\mathbb{I}}\nolimits} 
\newcommand{\A}{\mathop{\mathcal{A}}\nolimits} 
\newcommand{\sq}{\qquad $\blacksquare$} \newcommand{\wsq}{\qquad $\square$}

\newcommand{\NF}{\mathop{\mathcal{N}}\nolimits}
\newcommand{\Di}{\mathop{\mathrm{D}}\nolimits}

\newcommand{\X}{\mathop{\mathcal{X}}\nolimits}
\newcommand{\Y}{\mathop{\mathcal{Y}}\nolimits}

\newcommand{\num}{\mathop{\mathfrak{n}}\nolimits}
\newcommand{\mm}{\mathop{\mathfrak{m}}\nolimits}
\newcommand{\Capa}{\mathop{\mathfrak{C}}\nolimits}
\newcommand{\sig}{\mathop{\mathfrak{s}}\nolimits}
\newcommand{\rr}{\mathop{\mathfrak{r}}\nolimits}

\newcommand{\aff}{\mathop{\mathrm{aff}}\nolimits}
\newcommand{\co}{\mathop{\mathrm{co}}\nolimits}
\newcommand{\cone}{\mathop{\mathrm{cone}}\nolimits}
\newcommand{\ext}{\mathop{\mathrm{ext}}\nolimits}
\newcommand{\ri}{\mathop{\mathrm{ri}}\nolimits}
\newcommand{\Span}{\mathop{\mathrm{span}}\nolimits}

\newtheorem{Thm}{Theorem} 
\newtheorem{Prop}{Proposition} 
\newtheorem{Lem}{Lemma} 
\newtheorem{Cor}{Corollary} 
\newtheorem{Def}{Definition} 
\newtheorem{Rem}{Remark} 
\newtheorem{Conj}{Conjecture}

\newcommand{\bra}[1]{\langle #1 |}
\newcommand{\ket}[1]{| #1 \rangle} 
\newcommand{\bracket}[2]{\langle #1 | #2 \rangle}
\newcommand{\ketbra}[2]{| #1 \rangle \langle #2 | }

\newcommand{\norm}[1]{\Vert #1 \Vert}
\newcommand{\nnorm}[1]{|\hspace{-0.2mm}|\hspace{-0.2mm}| #1 |\hspace{-0.2mm}|\hspace{-0.2mm}|}

\newcommand{\red}[1]{ {\color{red} \ #1 \ }}
\newcommand{\bl}[1]{{\color{blue} #1 }}
\newcommand{\re}[1]{{\color{red} #1 }}
\newcommand{\cy}[1]{{\color{cyan} #1 }}

\title{

Information-induced asymmetry of state space in view of general probabilistic theories 
}
\author{Keiji Matsumoto}
\affiliation{National Institute of Informatics, 2-1-2, Hitotsubashi, Chiyoda-ku, Tokyo 101-8430}
\email{keiji@nii.ac.jp}
\author{Gen Kimura}
\affiliation{Shibaura Institute of Technology, Saitama, 337-8570, Japan}
\email{gen@shibaura-it.ac.jp}
\date[]{April. 28, 2016}
\begin{abstract}
It is known that the high-dimensional quantum state space is notoriously complicated in contrast with the beautiful Bloch ball of the qubit. 
We examined the mechanism behind this fact in the frame work of general probabilistic theory (GPT), and found rather general quantitative relations
between the geometry of the state space and its information storing capability. The main result is the
information-asymmetry identity, which (up to the constant term) equates the {\it Minkowski measure of asymmetry} 
with the {\it information storability} which, in addition to its own operational meaning, serves as an upper bound to common information measures
such as semi-classical capacity. As a consequence, the asymmetry measure is lower-bounded by information storing capability of the state space,
so the increase in the latter enhances the former.  
Coming back to the quantum systems, the $d$-level state space cannot be symmetric ``because" it can store more than a single bit of information.
Also, the Holevo capacity of any quantum channel with point-symmetric image is at most a single bit.
In the course of the research, we applied Shannon theory to GPT, producing a couple of new results.
Also presented is a new geometrical proof of known upper bounds to information measures.
\end{abstract}


\maketitle

\section{Introduction and Summary}\label{sec:intro}
Influenced by the development of quantum information theory, the study of foundations of quantum mechanics from the operational view points has been one of the important trends.
An underlying motivation is derivation of quantum theory by operational principles directly testable in experiments, without presupposing 
wave functions, Hilbert spaces, operator algebras, etc..
While some of them are directly aiming at this goal
 \cite{ref:F,ref:C,ref:H,ref:Hal,ref:rev,ref:MM,ref:CDP,ref:CDP2,ref:KNI,ref:NK},
others have shed light on relations among physical and informational principles, through extension of information theory to more generalized framework than quantum theory \cite{ref:Barrett1,ref:Barrett,ref:BHK,ref:SW,ref:BE,ref:KNI10,ref:BBKM15,ref:KNI2,ref:KMI}.



Building upon these lines of researches, the present paper is aimed to understand geometric properties of the state space from informational view points.
Our main finding is ``information-induced" asymmetry of the state space:
{\it The state space with large information storing capability is necessarily highly asymmetric}
(Theorem \ref{thm:main}). 

Our starting point is the following observations.
Except for the beautiful Bloch ball of a qubit, the space of density operators of a quantum system
is notoriously complicated, and increasingly so as the
number $d$ of levels grows. 
To be quantitative, the ratio between the shortest and the longest distance (in Frobenius norm) between the center (the maximally mixed state) and the surface of the state space is $d-1$
(See e.g. \cite{ref:IK,ref:KimKos} and references therein).

Interestingly, if we let $d$ denote the number of a sample space, exactly the same holds in classical systems, which is $d-1$ dimensional simplex. 
As $d$ is the maximum number of perfectly distinguishable states in both cases, the above observation seems to indicate relations between information storing capability and geometry of a state space. 

For deeper investigations, however, sticking to these two theories of nature does not seem fruitful:
The quantities we are considering might only be in the same value as the truly relevant ones due to their specific structures.
Thus, a more general framework including reasonable variety of theories is desired, not to let `accidental' relations obscure essential ones.

Fortunately, the general probabilistic theory (GPT) \cite{ref:Mackey,ref:Araki,ref:Gudder1,ref:Ludwig,ref:OrderedVec,ref:PR} perfectly serves this aim.
This is a generalization of quantum theory, and the commonly used theoretical framework in the aforementioned recent studies on its foundations. 
As such, it is neither too generic nor too specific. 
In our context, it is specific enough to define `information',  
and generic enough to allow of any (convex) set as its state space.
Also, its representation on a vector space is unique only up to affine transforms.

In this more general settings, our characteristic of the state space geometry is clearly inappropriate, as 
it cannot be affinely invariant, being norm-dependent.  

Thus this quantity is replaced by the one with desired invariance --
the Minkowski measure $\mm$ \cite{ref:MS}, 
which is commonly used as a measure of deviation from point-symmetry in convex geometry. ($\mm \in [1,\infty)$ and $\mm = 1$ iff the set has point-symmetry.)  
If the system is quantum or classical, 
\begin{equation}\label{eq:mdclqu}
\mm = d-1
\end{equation}
in accordance with our preliminary observation (Sec.~\ref{sec:MS}). 
So the complexity of the state spaces in our discussion may well be identified with the asymmetry in the sense of $\mm$.

\begin{figure}[h]
\includegraphics[width=8.5cm]{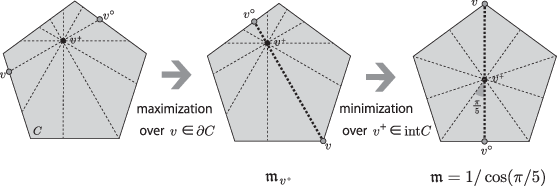}
\caption{
Minkowski measure for the regular pentagon. 
}\label{fig:pent}
\end{figure}

The relation \eqref{eq:mdclqu}, however, fails in many GP models,
if $d$ denotes the maximal number of perfectly distinguishable states.
For example, a regular-pentagon state space has $\mm = 1/\cos(\pi/5) \simeq 1.24$ (See Fig.~\ref{fig:pent}), which is not even an integer. 

Thus, to replace $d$, we introduce a new quantity, {\it information storability}, and denote it by $\num$.
This is the maximum of the `average number' of messages correctly stored by the information storing protocol in Sec.~\ref{sec:def-num},
a GPT version of a CQ channel \cite{ref:Holevo, ref:Holevo2, ref:HayashiNagaoka}. 
Besides its own operational meaning, $\num$ provides an upper bound of other commonly used information measures, such as the number of the perfectly distinguishable states and 
(the exponential of ) the semi-classical capacity, etc. 
\footnote{Also, $\num$ is related to max-relative entropy, which plays a significant role in quantum information theory \cite{ref:KRF,ref:Dmax}. See Appendix \ref{sec:pf-d<c<n}.}
(See \eqref{eq:d<c<n-F}, \eqref{eq:st-converse}, and \eqref{eq:l<n} in Sec.~\ref{subsec:info-measures}.).

Once these quantities are chosen, it is not hard to show the {\it information-asymmetry (IA) identity}
\begin{equation}\label{eq:mn}
\mm = \num - 1 
\end{equation}
for any GP model (Theorem \ref{thm:main} in Sec.~\ref{subsec:mk}).

This identity, combined with the aforementioned properties of $\num$, indicates that $\mm$ is equal to or larger than various 
measures of information storing capability.
Let us observe its consequences in quantum systems, for instance:
A qubit state space, being point symmetric ($\mm=1$), can store not more than a single bit ($\num=\mm+1=2$),
but a $d$-level system, capable of storing more bits, cannot be point-symmetric ($\mm=\num-1>1$).
So we may well say that this asymmetry of the state space is `{\it information-induced}'.
Needless to say, such `information-induced asymmetry' is observed in any GP systems.

As in many of researches of GPT, our results rely on the assumption that all the mathematically valid states and measurements are feasible.
But even without this assumption, a series of inequalities justifies our main message (Sec.~\ref{subsec:mk}, ~\ref{subsec:info-measures} and \ref{sec:CD}).

Another geometric factor related with information is the dimension of the state space:
See e.g. Proposition 6 of \cite{ref:KNI} and Theorem 2 of \cite{ref:FMPT}.
We show these are corollaries of our results and the known relation between $\mm$ and the dimension.
As the latter argument is purely geometrical, in view of relation between geometry and information,
the asymmetry is more essential than the dimension (Sec.~\ref{sec:degree}). 

At the end (Sec.~\ref{sec:csds}), a set of natural conditions sufficient for $\num=d$ is shown,
partly motivated by characterization of quantum theory.

To quantify information storing capability, we applied Shannon theory to GPT, as laid out in 
Sec.~\ref{subsec:info-measures} and Appendices~\ref{sec:pf-d<c<n}-\ref{sec:sig-dim}. 
These meant to be a first step towards the construction of GPT Shannon theory.

The proof of the IA-identity is an application of the strong duality of semidefinite program (SDP), 
a traditional tool in quantum information \cite{ref:Detection}, but not without non-trivial technicality
 (Appendix~\ref{sec:proof}).

Appendix~\ref{sec:rep} contains a construction of a vector space representation of GPT that 
uses the function giving probability rule rather than abstractly constructed `convex structure'\cite{ref:Gudder1},
to stress that a vector space representation is a mere rewriting of the probability rule.

The paper is organized as follows. 
After a brief review of the GPT (Sec.~\ref{sec:GPT}) and the Minkowski measure (Sec.~\ref{sec:MS}), 
the main body of the paper Sec.~\ref{sec:n} follows. 
Sec.~\ref{sec:def-num} introduces the information storability with a brief account of its information theoretic meanings,
whose detail will be in Sec.~\ref{subsec:info-measures}. 
Sec.~\ref{subsec:mk} is the full exposition of the IA-identity, our main result. 
The other results are presented after these subsections.
We conclude the paper in Sec.~\ref{sec:CD} with some discussion and an open problem.

\section{General Probabilistic Theories}\label{sec:GPT} 

This section is a brief review of GPT, a framework for probabilistic `operational' theories such as quantum theory.
See Appendix \ref{sec:rep} and \cite{ref:OrderedVec} etc. for details.

A GPT constitute at least of the set $\SA$ and $\M$ of states and measurements, 
and the function $\Pr[x|M,s]$ that gives the probability of observing the data $x$ for a state $s\in\SA$ of a system and a mesurement $M\in\M$.
Such a theory can be conveniently represented by a dual pair of ordered vector spaces. (Recall quantum theory is represented on the space of operators, which are vector spaces.) 

In the construction in Appendix~\ref{sec:rep}, we identify the state $s$ with the function $\Pr[\cdot|\cdot,s]=:\hat{s}$, 
and the measurement $M$ with the tuple $(e^M_x)_{x=1}^l$, where $e^M_x:=\Pr[x|M,\cdot]$. $e^M_x$ is called an {\it effect}, and the set of all the effects is denoted by $\E$. 

This representation does not distinguish two states if they are distinguished by no measurement, but such states need not be treated as distinct objects in explanation of observed data \footnote{See Appendix A and \cite{ref:Kraus}.}. So we identify the state $s\in\SA$ with $\hat{s}$, and by abusing the notation, we write $s$ instead of $\hat{s}$. By the same token, we identify $M\in\M$ with the tuple of the effects $(e^M_x)_{x=1}^l$.

Since $\hat{s}$'s and $e^M_x$'s are real valued functions, linear combinations and the order structures are naturally defined
Also, the pairing $\langle\cdot,\cdot\rangle$ is defined by bilinear extension of the relation 
\footnote{If $\SA$ and $\E$ in the beggining are given as  subsets of dual pair of ordered linear subspaces,  then the original addition, scalar multiplication. pairing, and order can be different from those defined in this manner. If this is the case, the following Born-like formula, the law of probabilistic mixture and so on may not be true with respect to the originally given structures, but it is not a contradiction. We only have to replace the structures by those defined in our recipe.}.  
\begin{equation}\label{eqn:prob}
\Pr[x|M,s]=\langle s ,e^M_x\rangle 
\end{equation}

Introduce norms and take closure of these spaces, and extend the pairing $\langle\cdot,\cdot\rangle$ and the order structures continuously. 
Then we obtain the dual pair $\B_*$ and  $\B(=(\B_*)^*)$, where the set $\SA$ and $\E$ is a subset of the former and the latter, respectively. 

In quantum theory, a state is a density operator, and $(e_x)_{x=1}^l$ is a POVM, and $\B_*$ and $\B$ are spaces of self- adjoint  operators\footnote{If the dimension of the underlying Hilbert space is infinite diensional, $\B_*$ and $\B$ is the space of self-adjoint trace-class operators and self-adjoint bounded operators.}, and $\langle A,B\rangle=\tr AB$, so \eqref{eqn:prob} corresponds to the Born rule.

The orders are introduced to $\B_*$ and $\B$ so that any states and effects are positive, and that the duality
\begin{align}
v\in \B_{*+} \Leftrightarrow \forall f\in\B_+ \, \langle v,f\rangle\ge 0  \label{eq:order-dual-1} \\
f \in \B_{+} \Leftrightarrow \forall v\in\B_{*+} \, \langle v,f\rangle\ge 0 \label{eq:order-dual-2}
\end{align}
holds. Here,  $\B_{*+}$ and $\B_+$ are the positive cones. Moreover, $\B_{*+}$ and $\B_+$ are norm-closed, pointed, and generating
\footnote{A  non-empty subset $V$ of a real vector space is a cone iff $v_1,v_2\in V\Rightarrow v_1+v_2\in V$ and $\lambda\ge 0, v\in V\Rightarrow \lambda v\in V$. 
It is pointed  iff $V\cap -V=\{0\}$. It is generating if $V+(-V)$ equals the whole vector space  \cite{ref:OrderedVec}. 
Some authors use the term `wedge' for our `cone', and use the term `cone' for our `pointed cone'. Also, some authors use the term `proper' for our `pointed'.
Also, there is some disagreement in the literature about the meaning of the term `proper cone' . See e.g., Remark 2.3, \cite{ref:OrdUnit}.
}.

Define the {\it unit} $u\in \B$, a generalization of the identity operator in quantum theory, by $\langle s,u\rangle=1 (\forall s\in\SA)$. 
($\langle v,u\rangle$ for an arbitrary $v\in\B_*$ is defined by extending it linearly and continuously. ) 
Then any effect $e\in\E$ should satisfy 
\footnote{$\sum_{x=1}^l e_x= u$ follows from $\sum_x\Pr[x|M,s]=\sum_x \langle s, e_x\rangle=1$. The inequality follows from that any effect is positive.}  
\begin{equation}\label{measurement}
 0\le e_x\le u,\: \sum_{x=1}^l e_x= u. 
\end{equation} 
 
The norm $\norm{\cdot}$ on $\B$ and $\norm{\cdot}_1$ are defined so that 
\begin{align}
\norm{f} &=\inf\{ \lambda ;-\lambda u\leq f \leq \lambda u,\, \lambda\ge 0 \ \}\label{eq:ordnorm}\\
&=\sup_{v\in \B_{*+}, \norm{v}_1=1} |\langle v, f\rangle|=\sup_{v\in \B_{*+}, \langle v, u\rangle=1} |\langle v, f\rangle| \label{eq:sup-norm}\\
&=\sup_{v\in \B_{*}, \norm{v}_1=1} |\langle v, f\rangle| \label{eq:sup-norm-2}\\
\norm{v}_1:&=\sup_{\norm{f}\le 1} |\langle v,f\rangle|,\label{eq:normE}
\end{align}
hold. In quantum theory, $u=\I$. Also $\norm{\cdot}$ and $\norm{\cdot}_1$ is the operator norm and the trace norm, respectively. 

Observe that $\mathrm{Pr}[x|M,s]$ is continuous
\footnote{Recall $|\langle v,f\rangle|\le \norm{v}_1\norm{f}$, which follows from \eqref{eq:sup-norm-2} or \eqref{eq:normE}. } : 
\begin{align}
&|\mathrm{Pr}[x|M,x]-\mathrm{Pr}[x|M',x]|\nonumber\\
&=|\langle s,e_x\rangle-\langle s',e'_x\rangle|\leq \norm{s-s'}_1+\norm{e_x-e'_x}. \label{eq:cont}
\end{align}
and that the identities 
\begin{align}
&\langle s, p e_x+q e'_x\rangle=p\mathrm{Pr}[x|M,s]+q\mathrm{Pr}[x|M',s],\nonumber\\
&\langle p s+q s', e_x \rangle=p\mathrm{Pr}[x|M,s]+q\mathrm{Pr}[x|M,s'],\label{eq:mixture}
\end{align}
follows from the definition of $\langle\cdot,\cdot\rangle$.   
The RHSs of \eqref{eq:mixture} are usually interpreted as the probabilistic mixture of the two experimental situations
\footnote{Note such an interpretation is {\it not} logically necessary, and requires an additional postulate. See Appendix~\ref{sec:rep} for the detail.}.  

For mathematical simplicity, unless otherwise mentioned, we suppose [R1] and [R2]: 

\noindent {\bf [R1]} $\dim \B<\infty$ 
\footnote{So $\B_*$ is identified with the dual $\B^*$ of $\B$. 
If $\dim \B=\infty$, 
this is not the case. This affects the dual representation of $\num$. Also, typically, $\num=\infty$, so some additional constraints on the states, e,g,, a constraint on `energy' may be needed as done in e.g., \cite{ref:Holevo2}.}.

\noindent {\bf [R2]} 
All the measurements with the effects 
corresponding to \eqref{measurement} is feasible, so $\E=\{e;e\in\B, 0\le e\le u\}$.
Also, $\SA$ is the following compact convex set: 
\begin{equation}\label{eq:S}
\SA = \left\{ s \in \B_{*+} \,;\, \left\langle s,u\right\rangle =1\right\}.
\end{equation}

There aren't compelling arguments for these assumptions. 
For example, in case of quantum systems, the law of dynamics may restrict physically feasible measurements and states. 
Meantime, these assumptions are commonly used, yielding fruitful results.
 In any case, {\it even without [R2], as long as $\SA$ is convex and compact, the essential part of our argument 
remains intact as will be expounded in Sec.~\ref{subsec:mk}.}

\medskip

A family of states $\{s_x\}_x$ is called 
{\it perfectly distinguishable} (shortly, {\it distinguishable}) 
if there is a measurement $M=(e_x)_x$ such that 
$\langle s_x, e_{x'}\rangle = \delta_{x,x'}$. 
Also, $d(\SA)$ denotes the maximal number of distinguishable states in $\SA$, abbreviated as $d$.
Unless $\SA$ is a singleton (which we won't treat as a trivial case), $d$ is always greater or equal to $2$ \footnote{This is seen by considering parallel supporting hyperplanes of $\SA$, which forms a measurement to distinguish the supporting two states \cite{ref:KMI}.}. We denote by $D$ the affine dimension of $\SA$: $D = {\rm dim}\SA = {\rm dim} \B_* -1$. 

\noindent{\bf [Quantum and Classical Systems] }
The state space $\SA_q$ for $d$-level quantum system is the set of density operators, so $D=d^2-1$. 
Meantime, The classical state space $\SA_{cl}$ with $d$-elementary events forms a $(d-1)$-dimensional simplex. 
The maximum number of perfectly distinguishable states is $d$. Also, $D = d-1$.	

\section{Minkowski Measure}\label{sec:MS}
Any state space $\SA$ and its image $A\SA$ by an invertible affine transform $A$ are ``equivalent".
Consider the linear extension of $A$ to $\B_*$, and denote it also by $A$. Replace the positive cone $\B_{*+}$ by $A\B_{*+}$.
Also replace $\E$ and $\B_+$ by $A'\E$ and $A'\B_+$, respectively, where $A'$ is defined by the relation $\langle Av, A' f\rangle=\langle v,  f\rangle (\forall v, f)$.
Clearly,  the pair of $A\SA$ and $A^*\E$ represents the same GPT as the pair $\SA$ and $\E$ does.

Therefore, affine invariance is a desired property of a characteristic of a GP state space. 
The Minkowski measure, widely used in convex geometry, provides a natural affine-invariant measure of point-asymmetry of a convex body \cite{ref:MS}. 

Let $C$ be a compact convex set in a finite dimensional vector space with the norm $\nnorm{\cdot}$.
We denote by $\mathrm{int}\,C$ and $\partial C$ the set of interior points and boundary points of $C$, respectively.
If $v_+\in \mathrm{int}\,C$, the {\it maximal distortion} $\mm_{v^+}$ with respect to $v^+ $ is  
\begin{equation}\label{eq:mo}
\mm_{v^+}(C) := \max_{v \in \partial C} \frac{\nnorm{v - v^+}}{\nnorm{v^\circ-v^+}}
\end{equation}
where $v^\circ$ is antipodal of $v$ about $v^+$, i.e., the other endpoint of $C$ from $v$ passing through $v^+$.  If $v_+\not\in \mathrm{int}\,C$, define $\mm_{v^+}(C) :=\infty$.
The {\it Minkowski measure} $\mm$ is defined by 
\footnote{Note also $\mm_{v^+}(C)to\infty$ as $v^+$ approaches the boundary from the inside.} 
\footnote{To authors knowledge, the first use of this measure in quantum information is by \cite{ref:GB}. They called $\mm$ ``coefficient of symmetry", being aware of existing theories on this quantity in convex geometry. Another precursor is \cite{ref:J}, where $(1+\mm_{v^+})^{-1}$ was independently introduced for some special cases.}. 
\begin{equation}\label{eq:m}
\mm(C) := \min_{v^+ \in C} \mm_{v^+}(C).
\end{equation}
Below, the dependency on $C$ is suppressed if not confusing. 
The set of minimizers of the RHS of \eqref{eq:m}, which is compact and convex, is called the {\it critical set} and denoted by $C^+$. 
Fig. 1 illustrates the maximal distortion $\mm_{v^+}$ and the Minkowski measure $\mm =1/\cos(\pi/5)$ for the regular pentagon $C$, where the critical
set is a singleton composed of the center point.

Minkowski measure has the following bound: 
\begin{equation}\label{eq:bound}
1 \le \mm \le {\rm dim}\, C -{\rm dim}\, C^+. 
\end{equation}
The lower and upper bound is attained iff $C$ is point-symmetric and a simplex, respectively. In either case, the critical set $C^+$ is a singleton \cite{ref:MS}.
 
$\mm_{v^+}$ and $\mm$ are invariant by affine transforms. 
To see this, let us write them without resorting to norms.
If  $v^+\in\mathrm{int}C$, $v^\circ -v^+=-a_{v^+,v}(v -v^+)$
for some $a_{v^+,v}\ge 0$, which is decided by
\begin{align}
a_{v^+,v} = \max \{a \ ; \ v^+-a(v -v^+) \in C,\, a\ge 0 \}. \label{eq:a}
\end{align}
This is clearly affinely invariant, and equals $\nnorm{v^\circ-v^+}/\nnorm{v - v^+}$. So by \eqref{eq:m}, .
\begin{align}
\mm_{v^+} &= \max_{v \in \partial C} (a_{v^+,v})^{-1}\nonumber\\
               &=\max_{v \in \partial C} \min\{c \ ; \ \frac{1}{c-1} (cv^+ - v) \in C, c\ge 1\} \label{eq:ma}
\end{align}
where the second identity is by the change of parameter to $c:=a^{-1}+1$, and this quantity is affine invariant. 
If $v^+\in\partial C$, this last end is clearly $\infty$.

When $C=\SA$, \eqref{eq:ma} can be rewritten as 
\begin{align}
\mm_{s_0} +1 &\underset{\mathrm{(i)}}{=} \max_{s\in\partial\SA}\min\{ c \ ; \ cs_0 \ge s \} \nonumber\\
        &= \min\{ c \ ; \ \forall s\in\partial\SA \ cs_0 \ge s \}.\nonumber\\
       &\underset{\mathrm{(ii)}}{=} \min\{ c \ ; \ \forall s\in\SA \ cs_0 \ge s \}, \label{eq:mm-gp}
\end{align}
where the identity (i) is by
\begin{align}
cs_0- s\ge 0 \Leftrightarrow \exists c'\ge 0 \ c'(cs_0-s)\in \SA \label{eq:sym-ord}.
\end{align}
If  $s_0\in\partial\SA$, there is no $c'$ satisfying the condition 
\footnote{Otherwise, there is $s_1\in \SA$ with $c'(cs_0-s)=s_1$, so $s_0=(1/c)(s+c's_1)$. But this cannot be the case if  $s_0$ is on the boundary and $s$ is in the interior.},
and the RHS of the first identity is $\infty$. The identity (ii) is by convexity of the positive cone. 

For another interesting and useful expression, see Appendix~\ref{sec:mm-another}.

\medskip

\noindent{\bf [Quantum and Classical Systems]} If $\ket{\phi}$ is the eigenvector of $\rho_0$ for the minimum eigenvalue $\lambda_{\min}$,
\begin{align*}
\mm_{\rho_0}+1&\ge \min\{ c \ ; cs_0 \ge \ket{\phi}\bra{\phi} \}=(\lambda_{\min})^{-1}, \\
\mm+1&\ge \max_{\rho_0\in\mathrm{int}\SA}(\lambda_{\min})^{-1}=d. 
\end{align*}
Meantime,
\begin{align*}
\mm+1\le \mm_{\I/d}+1=\min\{ c \ ; \forall s\in\SA \ cd^{-1}\I \ge s \}=d. 
\end{align*}
Therefore, $\mm=d-1$, recovering \eqref{eq:mdclqu}.
(In fact , $\mm_{\rho_0}=(\lambda_{\min})^{-1}-1$ as is shown in Appendix~\ref{sec:mm-another}.) 
The critical set is a singleton composed of the maximally mixed state, so $\dim \SA_q^{+}=0$.

Similarly, in classical systems, $\mm=d-1$ and $\dim \SA_{cl}^{+}=0$.  
Also, $d= D+1$, and the upper bound \eqref{eq:bound} is attained.
These can be confirmed also by noticing that $\SA_{cl}$ is a simplex (See Example 2.1.7 in \cite{ref:MS}.).
	
\medskip

Summarizing, in both the classical and quantum state spaces $\mm=d-1$. 
However, as mentioned in the introduction, \eqref{eq:mdclqu} fails to hold in general: 

\medskip 

\noindent{\bf [Regular-pentagon model]} Let $\B=\B_*:=\R^3$ and $u:=(0,0,1)$. Let $\SA$ be a regular-pentagon on a 2-$\dim$ plane $z=1$, centered at $(0,0,1)$. By elementary geometry, 
$\mm=1/\cos (\pi/5)\simeq 1.24$. This cannot equal $d-1$, not being an integer.
In fact, $d=2$ in this model
\footnote{ By \eqref{eq:d<c<n<D} or Proposition 6 in \cite{ref:KNI}, $d \le D + 1$ and ``=" holds iff the system is classical. The assertion follows as $D=2$ and $\SA$ is not a simplex.}. 

\medskip

To establish a general relation between state space geometry and information, we will introduce the {\it information storability} in the next section. 

\section{Information Storability}\label{sec:n} 
\subsection{Definition and information theoretic meanings}\label{sec:def-num}
In this paper, the term `information' is always associated with the number of classical messages which can be stored into the system:
A classical message $x$ $(=1,\cdots,l)$ is encoded into a state $s(x)$ in a set $F\subset\SA$ of states, and 
the decoding process is represented by a measurement $M=(e_x)_{x=1}^l$ on $s(x)$.

It is a GPT version of  CQ (classical-to-quantum) channels, widely used mathematical model of classical message sending by quantum channels \cite{ref:Holevo, ref:Holevo2, ref:HayashiNagaoka}.
Here, $s(x)$ is the state which is the degraded by noisy channels, and the receiver decodes the message $x$ by measuring  it. 
In general, $s(x)$ cannot be an arbitrary state, but is an element of a subset $F$ of $\SA$.
The celebrated Holevo-Schumacher-Westmorand channel capacity formula was proved in an asymptotic version of this setting.  

Below, mostly we treat noiseless channels, where $F=\SA$. 
Such a problem is trivial in the case of quantum or classical systems, but it is not the case for other GPTs.  

To discuss the trade-off between the number of messages $l$ and the success probability of decoding 
\begin{align}
 P_{suc}(s(\cdot),M):=\frac{1}{l}\sum_{x=1}^l\langle s(x),e_x \rangle, \label{eq:p-suc}
\end{align} 
let us consider the maximization of the product of them:
\begin{align}\label{primal}
\num(F):=\sup_{s(\cdot),M,l } l\cdot P_{suc}=\sup_{s(\cdot),M,l } \sum_{x=1}^l\langle s(x),e_x \rangle,
\end{align}
where, $s(x)$ ($x=1,\cdots,l$), $M$, and $l$ runs for all the states in $F$, all the measurements that take values in $\{1,\cdots,l\}$, and all the natural numbers, respectively. 
We call $\num(F)$ the {\it information storability} of $F$. 
If not confusing, $\num(\SA)$ is simply denoted by $\num$. 

We rewrite \eqref{primal} to its dual form: 
\begin{eqnarray}
\num
=\min_{s_0 \ge 0 \langle s_0,u\rangle=1}\left[\min \left\{ c\,;\,\forall s \in \SA\, c\,s_0\ge s \right\}\right].  \label{N-ratio}
\end{eqnarray} 
The proof uses a traditional technique of signal detection of quantum information \cite{ref:Detection, ref:Detection-2}, but not without a difference from existing similar results.
See Appendix \ref{sec:proof} for the proof and the detail.

For illustration and the later use, without using [R2], we show 
\begin{eqnarray}
\num\le\min_{s_0 \ge 0 \langle s_0,u\rangle=1}\left[\min \left\{ c\,;\,\forall s \in \SA\, c\,s_0\ge s \right\}\right].  \label{eq:weak-dual}
\end{eqnarray}%
If $cs_0\ge s$ for all $s\in F$, each $s(x)$ in the last end of \eqref{primal} is upper-bounded by $c s_0$. 
So $c\ge\num$, and the minimization over $c$ leads to the desired inequality.

Information storability $\num(F)$, not only being the maximum of this figure of the merit, 
can be used for evaluation of the trade-off,
\begin{align}
 \log_2 l\le \log_2 \num(F) -\log_2 P_{suc}. \label{eq:l<n}
\end{align}
If $P_{suc}=1$, the RHS is $\log_2 d$, so $\num$ is an upper bound to $d$.
Also, if $P_{suc}$ is not too small, in many cases $\log_2 \num$ is dominant in the RHS, giving a reasonable upper bound of $\log_2 l$.

In Sec. \ref{subsec:info-measures}, we discuss the relation of this quantity with other information theoretic quantities such as the Shannon and the Holevo capacity, 
and show $\num$ is an upperbound to most of them. Here, we point out that $\num$ is closely related to {\it max-relative entropy}
\begin{equation}\label{eq:dmax}
\Di_{\max}(s_1 \Vert s_2):=\min \{\lambda ;s_1 \leq 2^{\lambda }s_2\}, 
\end{equation}
which plays significant role in quantum information theory \cite{ref:KRF,ref:Dmax}.
By \eqref{N-ratio},
\begin{equation}
\log_2 \num=\min_{s_0 \in \mathcal{S}}\max_{s \in \SA}\,\Di_{\max}(s \Vert s_0). 
 \label{N-ent}
\end{equation}%
 
\subsection{Information-asymmetry identity} \label{subsec:mk}

\begin{Thm}\label{thm:main} (Information Asymmetry (IA) identity) For any GP model, 
\begin{equation}\label{eq:main}
\num = \mm + 1. 
\end{equation}
\end{Thm} 
\noindent[Proof] The statement of the theorem is obtained by combining  \eqref{eq:m}, \eqref{eq:mm-gp} and \eqref{N-ratio}. \hfill$\blacksquare$ 
 
\medskip 

Theorem~\ref{thm:main} establishes a link between information theory and geometry of the sate space,
by equating the asymmetry measure with an information theoretic quantity $\num$,
which is operationally defined by \eqref{eq:p-suc}. As will be expounded in Sec.~\ref{subsec:info-measures}, 
$\num$ as an upper bound to various information measures. 
  
So IA identity indicates that the information storage requires asymmetry: If the state space can store may bits, that state space should be highly asymmetric. 
{\it Also, if the state space is point-symmetric, it can capable of storing only a single bit, and those capable of storing more are necessarily asymmetric.}

Coming back to the quantum theory, this `explains' the reason why only a qubit is allowed of a point-symmetric state space. 

Importantly, the above message does not rest on [R2] in Sec.~\ref{sec:GPT}, feasibility of all the mathematically valid states and measurements:
In the remainder of this subsection, we do not assume [R2], but we still suppose  $\SA$ is compact and convex. 
Even in this setting, it holds that
\begin{align}
   \num &\le \min_{s_0 \in \mathcal{S}}\max_{s \in \SA}\,\Di_{\max}(s \Vert s_0) \label{eq:w-IA-3}\\
   &\le \mm+1 \label{eq:w-IA}
\end{align}

\noindent [Proof of \eqref{eq:w-IA-3},\eqref{eq:w-IA}] 
Since \eqref{eq:weak-dual} is still intact in the present setting, \eqref{eq:w-IA-3} holds. 
Also,  $\Leftarrow$-part of \eqref{eq:sym-ord} is still intact, so is $\ge$-part of \eqref{eq:mm-gp}. Combined with \eqref{eq:m}, 
we obtain \eqref{eq:w-IA}.\hfill$\blacksquare$

In addition, information theoretic meanings of $\num$ and $\mm$ demonstrated in Subsec. \ref{subsec:info-measures} does not rest on [R2]. 

A restriction of the set of the states to $F\subset \SA$ may be viewed as a GP system with the state space $F$ and the measurements $\M$. 
This may be viewed as a model of, e.g., the noisy channel, where $F$ is the set of all the signal states reaching to the receiver's system,
or other humanitarian restriction of feasible states.
The asymmetry measure $\mm(F)$ of $F$ may be larger than it of $\SA$, but clearly, its information storing capability can only be smaller.
This is not an inconsistency at all, as $\mm(F)$ only sets an upper bound. 

On the other hand, if $\mm(F)$ is strictly smaller than $\mm(\SA)$, it gives a better bound on the information storing capabilities. 
As the uppr bound is of interest, $F$ may be replaced by any larger  compact convex set of states:
\begin{align}\label{eq:w-IA-2}
   \num(F)\le \inf_{F\subset F'\subset \SA}\mm(F')+1,
\end{align}
where $F'$ is compact and convex. This inequality implies a refined version of the `information induced asymmetry':
{\it If a set of the states $F$ can store more than a single bit reliably, $F$, nor any set of states $F'$ containing $F$, cannot be point symmetric.}

Clearly, this geometric statement about channels applies to quantum and classical systems.
For example, combined with \eqref{eq:st-converse-2} in the next subsection, this inequality shows:
if $F$ is the image of a quantum channel and if it is a subset of point-symmetric set of states, 
the Holevo capacity (the maximal rate of bits reliably transmitted by
separable signal states and collective measurements) cannot be more than a single bit.

\subsection{Relation to channel capacity}\label{subsec:info-measures}

{\it Different from other parts of the paper, in this subsection, [R2] is not assumed.} 

Besides its own operational meaning, information storability serves as an upper bound of performance measures of asymptotic versions of the information storing protocol.
Combined with IA identity \eqref{eq:main}, each of these inequalities adds another information theoretic meaning to the asymmetry measure $\mm$.

Channel capacities in GPT  may be defined by mathematical/formal analogy, via Shannon entropy. 
But such mathematical analogy may miss operational meanings of the channel capacity in information theory.
Therefore, we define them in purely operational manner, via the analysis of asymptotic version of information storing,
where where messages are stored in $n$ (to be taken to $\infty$) of parallely and separately prepared identical GP systems. 
It is a GP analogue of the theory of CQ channels, which led to introduction of the Holevo capacity
 and the celebrated Holevo-Schumacher-Westmoreland theorem\cite{ref:Wilde}. 
The argument below therefore runs in parallel with theory of QC channels. 

First, we discuss the case where no non-classical correlation is used both in encoding and decoding
\footnote{See e.g.~\cite{ref:FN, ref:Wilde} for its quantum version. They represent a channel as a map of a state space to another, 
but as commented above, the channel can be incorporated into the definition of the signal states.}.

In the encoding, a message $z\,(=1,\cdots,l_n)$ is first encoded to a tuple $x^n(z)=(x_1(z),\cdots,x_n(z))$ of symbols in $\X$, and then 
each $x_i(z)$ is encoded to a state $s_i(x_i(z))$ ($i=1,\cdots,n$). 
To decode the message, measure each $s(x_i(z))$ by $M_i=(e_{i,y})_{y\in\Y}$, and from the tuple of the results $y^n=(y_1,\cdots,y_n)$, compute an estimate $\psi^n(y^n)$ of $z$. 
Here, the $i$-th measurement $M_i$ may depend on the preceding measurement results $y_1,\cdots,y_{i-1}$.

The {\it semi-classical capacity} $\Capa(\SA)$, or simply $\Capa$, is the maximal rate of bits reliably encoded in such a setting:
\begin{align}
\Capa(\SA):=\sup\{\varliminf_{n\to\infty} \frac{1}{n}\log_2 l_n\, ;\lim_{n\to\infty}P^n_{suc}=1\}, 
\label{eq:rate}
\end{align}
where $l_n$ is the number of messages, and $P^n_{suc}$ is the success probability of decode,
and the supremum is taken over all the protocols expounded above. $\X$ and $\Y$ are arbitrary finite sets, and their sizes are to be optimized as well.

For such protocols, it is known that 
\cite{ref:FN,ref:Wilde}\footnote{The use of the randomness is not considered here, as obviously useless:The maximum of the success probability should be achieved at the extreme points. \eqref{eq:defC} holds also for a larger class of protocols, where adaptive choice of the measurements is allowed, see \cite{ref:FN}:
Their proof written for quantum systems easily generalizes to GPTs.
Coming back to our setting, $\le$ will be proved in the main text. 
$\ge$ can be proved as follows. First, fixed the sequence $\{s_i(\cdot),M_i\}_{i=1}^{\infty}$. Then the problem reduces to the channel coding of classical non-identical memoryless channels. So classical techniques of the information theory lead to that the rate more than $\varlimsup_{n\to\infty}\frac{1}{n}\sum_{i=1}^n C_{sh}(s_i(\cdot),M_i)$ is impossible. In particular, by the Arimoto bound \cite{ref:Arimoto}, we can prove this statement even under weaker constraint on the success probability, $\varliminf_{n\to\infty}P_{suc}^n>0$.
So, maximization about $\{s_i(\cdot),M_i\}_{i=1}^{\infty}$ leads to the assertion. },
\begin{equation}\label{eq:defC}
\Capa=\sup_{\X,\Y}\max_{s(\cdot),M} \Capa_{sh}(s(\cdot),M),
\end{equation} 
where $\Capa_{sh}(s(\cdot),M)$ is the the Shannon capacity \cite{ref:CT}
of classical channel $P[y|x]=\langle s(x),e_y\rangle$:
\begin{equation}
\Capa_{sh}(s(\cdot),M)= \inf_{p_0} \sup_{x\in\X} \Di(p^M_{s(x)}\Vert p_0),
\label{eq:C-formula}
\end{equation} 
where $p^M_{s(x)}(y):=\langle s(x),e_y\rangle$ and $p_0$ runs over all the probability distributions on $\X$, 
and $\Di(p\Vert q):=\sum_x p(x) \log_2 (p(x)/q(x))$ is the relative entropy
\footnote{This representation appears in \cite{ref:Ciszsar, ref:CT,ref:OPW, ref:SchWest}. 
It is derived by the min-max theorem from more common representation that uses mutual information .}
\footnote{In \cite{ref:FMPT}, they call the r.h.s of \eqref{eq:defC} ``Holevo capacity", but we reserve this term for the capacity of a quantum channel with collective measurements \cite{ref:Wilde}.}.

For illustration, we show that the rate given by the RHS of \eqref{eq:defC} is achieved.
Suppose $s_1=\cdots=s_n=s$ and $M_1=\cdots=M_n=M$. Then 
$x^n(\cdot)$ and $\psi^n(\cdot)$ plays the role of the encoder and the decoder of the $n$-times use of the classical channel $P[y|x]=\langle s(x),e_y\rangle$, respectively. 
So by Shannon's channel coding theorem, the maximal rate for each fixed $s(\cdot)$ and $M$ is $\Capa_{sh}(s(\cdot),M)$.

\begin{Thm}\label{thm:d<c<n}
\begin{align}
\log_2 d &\le \Capa  \le \log_2 \num. \label{eq:d<c<n-F}
\end{align}
\end{Thm}

As we saw in Sec.~\ref{sec:MS}, $\num=\mm+1=d$ in classical and quantum systems. So applying $\SA_q$ or $\SA_{cl}$ for $\SA$, all the inequalities in \eqref{eq:d<c<n-F} saturate. 
But some GP models  (e.g., the regular-pentagon model) do not have this property. For a sufficient condition for this saturation, see Sec.~\ref{sec:csds}.

In Appendix~\ref{sec:pf-d<c<n}, we show a proof of the relation which is shorter but uses the assumption [R2].

\noindent [Proof]
The first inequality is clear as $\log_2 d$ is the optimal rate under $P^n_{suc}=1\,(\forall n)$, 
so it remains to show the second one.

Consider the family $F_i:=\left\{P_i[\cdot|x]\right\}  _{x\in\X}$ of probability distributions,
where $P_i[y|x]:=\langle s_i(x),e_{i,y}\rangle$, and $s_i(x)$ and $M_i=(e_{i,y})_{y\in\Y}$ is a signal state and a measurement on the $i$-th system.
Here, the measurement $M_i$ ($P_i[\cdot|x]$ and $F_i$ as well) may depend on the outcomes $y^{i-1}:=(y_1,\cdots,y_{i-1})$
of the preceding measurements $M_1,\cdots,M_{i-1}$. If necessary,  we write, e.g., $P_i[\cdot|x,y^{i-1}]$ to indicate this dependency.
Consider also the set of the joint distributions
\begin{align*}
F^{(n)}:=\left\{  P^n[\cdot|x^n,]\right\} _{x^n\in\X^n}=\left\{ \Pi_{i=1}^nP_i[\cdot|x^n]\right\}_{x^n\in\X^n}.
\end{align*}
$F_i$'s and $F^{(n)}$ can be regarded as subsets of classical states.

Denote by $\num(F_i)$ and $\num(F^{(n)})$ the information
storability of the former and the latter:
\begin{align}
\num(F_i)=\sum_x\langle s_i(x),e_{i,x}\rangle
\end{align}
Taking maximum of  for the signal states and measurements,  we obtain (Recall \eqref{eq:p-suc}): 
\begin{equation}
\num=\max\,\num(F_i)\label{n-max}.
\end{equation}

Also, as classical systems satisfies [R2], we can use the duality \eqref{N-ratio} to compute
$\num(F_i)$ and $\num(F^{(n)})$:
\begin{align}
\num(F_i)  & =\min\{\,\sum_{y\in\Y}q(y);\,q(y)
\geq \max_{x\in\X}P_i[y|x,,y^{i-1}]\}\nonumber\\
& =\sum_{y\in\Y}\max_{x\in\X}P_i[y|x,y^{i-1}],\label{n-cl}
\end{align}
and
\begin{align*}
&\num(F^{(n)})  
  \underset{\mathrm{(i)}}{=}\sum_{y^n\in\Y^{\times n}}\max_{x^n\in\X^{\times n}} P^n[y^n|x^n] \\
&=\sum_{y^{n-1}}\sum_{y_n}\max_{x}P[y_n|x,y^{n-1}] \max_{x^{n-1}} P^{n-1}[y^{n-1}|x^{n-1}]\\
&\underset{\mathrm{(ii)}}{\le} \num \sum_{y^{n-1}} \max_{x^{n-1}} P^{n-1}[y^{n-1}|x^{n-1}]
... \le\num^n.
\end{align*}
where $(i)$ is by \eqref{n-cl} and 
$(ii)$ is by \eqref{n-max}.

Therefore, using \eqref{eq:l<n}, we obtain, to each fixed tuple of signal states and the
measurements,
\begin{align}
\frac{1}{n}\log_{2}\,l^n  
& \leq\frac{1}{n}\log_{2}\num(F^{(n)})-\frac{1}{n}\log_{2}\,P_{suc}^n \nonumber\\
& \leq\log_{2}\mathfrak{n-}\frac{1}{n}\log_{2}\,P_{suc}^n. \label{eq:l-n-p}
\end{align}
Taking the limits and the maximum, we obtain the asserted inequality.
\hfill$\blacksquare$

The above analysis is easily extended to the optimal rate under more relaxed constraint  
$\varlimsup_{n\to\infty} P_{suc}^n>0$
which allows non-negligible decoding error (`strong converse')
\footnote{Such a constraint is considered in discussing `strong converse' theorems. See e.g., \cite{ref:Arimoto, ref:strong-converse}.}.
Even in such a case, the rate is bounded as
\footnote{In fact, the optimal rate under this constraint equals the RHS of \eqref{eq:defC} as well. See the footnote right before \eqref{eq:defC}.} 
\begin{align}
 \varlimsup_{n\to\infty} P_{suc}^n>0\Rightarrow \varliminf_{n\to\infty} \frac{1}{n}\log_2 l_n \le \log_2 \num, \label{eq:st-converse}
\end{align}
as the second term of \eqref{eq:l-n-p} vanishes also under this condition. 

So far, we had not considered physical interactions between GP systems.
However, in the setting where the HSW theorem was shown, they assume  decoding process exploiting the interactions, 
while encoding is done in the same manner in our preceding analysis\cite{ref:Wilde}. 

Below, we discuss GP analogue of this setting, supposing that any measurement should be represented by a tuple of multi-affine functionals $(e^n_z)_z$ on $n$-tuple of states.
(But we are {\it not} arguing all such measurements are physically feasible.).
Here $e^n_z$ is positive on $\SA^{\times n}$,  and $\sum_z e^n_z =u$, where $u^n(s_1,\cdots,s_n)=1$ for all $(s_1,\cdots,s_n)\in\SA^{\times n}$.
They are linearly extended to $\B_*^{\otimes n}$, and can be viewed as an element of $(\B_*^{\otimes n})^*=\B^{\otimes n}$.
So we write  $e^n_z(s_1,s_2,\cdots,s_n)=\langle e^n_z, \otimes_{i=1}^n s_i\rangle$. 
Also, we denote the closure of the convex hull (the set of all the probabilistic mixtures) of 
$\SA^{\times n}$ by $\SA^n$$(n=1,2,\cdots)$
\footnote{Here we take closure with respect to the norm topology. 
Since $\dim\B$ is finite, all norm topologies are equivalent.}. 
Note $\SA^1$ is the closure of the convex hull of $\SA$.  

From here to the end of the subsection, $\B^n_{*+}$$(n\ge 1)$, the positive cone of $\B_*^{\otimes n}$ is defined as the cone generated by $\SA^n$. 
Then $\SA^n$ is the set of positive elements with $\langle v^n,u\rangle=1$, $v^n\in\B_*^{\otimes n}$. 
Also, \eqref{eq:S} and \eqref{eq:mm-gp} are valid, if $\min$ are replaced by $\inf$ whenever necessary. 

Such a composite GP system is called `minimal tensor product' \cite{ref:Hardy2}. 
But here we do not postulate that all the positive functionals on $\SA^n$ are physically feasible
nor that restriction of the state space to $\SA^n$ is due to the fundamental physical law. 

We encode the message $z\in\{1,...,l_{n}\}$ into $n$-tuple of states $s_{1}(z),\cdots,s_{n}(z)$, where $s_i(z)\in \SA$.
Here we don't consider probabilistic mixtures of $n$-tuples of states, as being obviously useless
\footnote{Recall the constraint of the optimization is affine in the signals. 
Therefore, even if probabilistic mixtures of the tuple of states $(s_1,\cdots,s_n)$ can be used for signals, optimal success probability of decoding is achived by
extreme points. }.

In this setting, we have 
\begin{Thm}
\begin{align}
 \varlimsup_{n\to\infty} P_{suc}^n>0\Rightarrow \varliminf_{n\to\infty} \frac{1}{n}\log_2 l_n \le \log_2 (\mm+1). \label{eq:st-converse-2}
\end{align}
\end{Thm}

\noindent [Proof] By \eqref{eq:mm-gp}, let $s_0$ be a state with $cs_0\geq s$ for any $s\in \SA$, where $c=\mm+1$.
Then by the positivity of $e^n_z$, we have $\langle (cs_0 -s_1)\otimes s_2\cdots, \otimes s_n, e^n_z\rangle \ge 0$, or 
$$c \langle s_0\otimes s_2\cdots, \otimes s_n, e^n_z\rangle\ge \langle s_1\otimes s_2\cdots, \otimes s_n, e^n_z\rangle$$
 for any $s_i\in \SA$, $i=1,\cdots,n$. Recursively, we obtain 
\begin{align}
c^n \langle s_0\otimes\cdots \otimes s_0,e^n_z \rangle \ge \langle s_1\otimes\cdots \otimes s_n, e^n_z \rangle\label{eq:min-mul}
\end{align}
Therefore, 
\begin{align*}
P_{suc}^{n}&=\frac{1}{l_{n}}\sum_{z} \langle s_1(z)\otimes\cdots  s_n(z), e^n_z \rangle
\leq\frac{c^{n}}{l_{n}}\sum_{z} \langle s_0^{\otimes n}, e^n_z \rangle\\
&=c^{n}/l_{n}.
\end{align*}
By moving terms and taking the limit, we obtain the desired inequality.\hfill$\blacksquare$

In the end, let us discuss if these upper bounds are reasonablly close to the optimal. 
There is no strong case for it, since we could not derive no non-trivial achievable rate.

But our upper bound $\num$ or $\mm+1$ is related to $\Di_{\max}$ by \eqref{N-ent}.
In the case of the quantum system, `$\epsilon$-smoothed' version of $\Di_{\max}$, 
after taking appropriate limits,  leads to the Holevo bound \cite{ref:Dmax,ref:KRF,ref:T}.

Therefore we discuss the $\epsilon$-smoothed version of $\mm$. Below, we denote $\mm(\SA^n)$ by $\mm^n$ and so on: 
\begin{align*}
&\mm_{s_0}^{\epsilon,n}+1\\
&:=\inf\{c; \forall s\in \SA^n\exists s'\ge 0 \norm{s-s'}_1\le \epsilon,\, cs_0\ge s' \}.\\
&\mm^{\epsilon,n}:=\min_{s_0\in\SA^n}\mm_{s_0}^{\epsilon,n}\le \mm^n.
\end{align*}
This $\mm^{\epsilon,n}$ is related to $\epsilon$-smoothed version of $\Di_{\max}$ by the relation analogous to \eqref{N-ent}
\footnote{The $\epsilon$-smoothed version of  $\Di_{\max}$ is defined by 
$\Di_{\max}^\epsilon(s_1\Vert s_0):=\inf_{s\ge 0, \norm{s-s_1}_1\le\epsilon}\Di_{\max}(s\Vert s_0)$. 
$\mm^{\epsilon,n}+1$ equals 
$\inf_{s_0\in\SA^n}\max_{s_1\in \SA^n}\Di_{\max}^\epsilon(s_1\Vert s_0)$.
}. 
For any $\epsilon>0$,
\begin{align*}
 &P_{suc}^n =l_n^{-1}\sum_{x=1}^{l_n}\langle z(x),e_z^n\rangle \nonumber\\
  &\le l_n^{-1}\sum_{z=1}^{l_n}\langle s'(z),m_x\rangle+\norm{s'(z)-s(z)}_1  \nonumber\\
  &\le l_n^{-1}\sum_{z=1}^{l_n} \langle( \mm^{\epsilon,n}+1)s_0,m_x\rangle+\epsilon = l_n^{-1} ( \mm^{\epsilon,n}+1)+\epsilon
\end{align*} 
Therefore, Since  $\mm^{\epsilon,n}$ is not larger than $\mm^n$, it may possibly give an improved upper bound to $P_{suc}^n$.
Indeed, in case of quantum system, we obtain the Holevo bound by taking the limit $n\to\infty$ and then $\epsilon\to 0$.  

However, recall we are working on the cases corresponding to noiseless channels.. 
In case of the quantum noiseless channel, the $\epsilon$-smoothing does not improve the bound. 
We show this also the case in any GPT, and argue that the bound by $\mm+1$ 
might be a reasonable upper bound to the rate of the messages storable in GP systems
\footnote{We realize this `evidence' is not very persuasive, but still better than none.}.

Below, we show 
\begin{align}
\lim_{\epsilon\downarrow 0} \varlimsup_{n\to\infty}\frac{1}{n}\log_2 (\mm^{\epsilon,n}+1)= \varlimsup_{n\to\infty}\frac{1}{n}\log_2 (\mm^n+1).\label{eq:ne=n}
\end{align}

Clearly, we only have to show `$\ge$'. Consider $s_0,\,s_1$ with 
\begin{align}
&\forall s\in\SA^n\,\, (\mm^n+1)s_0\ge s \nonumber\\
&\forall s\in\SA^n \exists s'\ge 0\, \norm{s-s'}_1 \le \epsilon, s'\le (\mm^{\epsilon,n}+1+\delta) s_1 \nonumber 
\end{align}
for an arbitrary $\delta>0$. 
Then 
\begin{align*}
s
&=s'+(s-s')\le s'+(s-s')_+ \\
&\le (\mm^{\epsilon,n}+1 +\delta)s_1 + \epsilon\, (\mm^n+1)s_0\\
&=\{(\mm^{\epsilon,n}+1 +\delta)+ \epsilon (\mm^n+1)\}s_2,
\end{align*}
where $(\cdot)_+$ is the positive part\footnote{See Appendix \ref{.sec:rep}} and $2\langle (s-s')_+,u\rangle=\norm{s-s'}_1\le\epsilon$. 
Also, $s_2$ is an appropriately defined state.
So, $\mm^n+1\le (\mm^{\epsilon,n}+1+\delta) + \epsilon( \mm^n+1)$. 
As $\delta>0$ is arbitrary, 
\begin{align*}
\mm^{\epsilon,n}+1 \ge (1-\epsilon)(\mm^n+1).
\end{align*}
Thus we obtain `$\ge$'. Since `$\le$' is trivial, \eqref{eq:ne=n} is proved.

Recently, another information measure, {\it signaling dimension}, is defined\cite{ref:FrenkelWeiner}, and already applied to GPT \cite{ref:DBTBV}. This new measure quantifies efficiency of simulation of a given physical system by a classical channel
\footnote{This is reminiscent of ``reverse Shannon"-type theorems \cite{ref:Wilde}, that treat asymptotic simulation of quantum channels by classical channels with assistance of pre-shared entanglement, though \cite{ref:FrenkelWeiner} is non-asymptotic and uses only shared randomness. },
 while other measures, including $\num$, concern information storage. Thus, it is expected (and easily proved) that this new one is an upper bound of the conventional measures.
See Appendix~\ref{sec:sig-dim} for the detail.

\subsection{Dimension and information}\label{sec:degree}
Clearly, the affine dimension $D$ of the state space sets an upper bound to the information storing capability:
Proposition 6 in \cite{ref:KNI} and Theorem 2 in \cite{ref:FMPT} had showed the tight upper bound of $d$ and $\Capa$ in terms of $D$
\footnote{As noted already, their definition of $\Capa$ is \eqref{eq:defC}, or the maximization of the measured mutual information, and they call it `Holevo capacity'.}, 
respectively.
They also had showed their bounds saturate iff the system is classical.

These are in fact corollaries of our results: 
By \eqref{eq:d<c<n-F} and the IA idnentity \eqref{eq:main}, the problem reduces to the relation between $D$ and $\mm$.
So by \eqref{eq:bound} and the condition for $\mm=D$, we obtain:
\begin{Thm} For any GP model with $D = {\rm dim} \SA$, 
\begin{equation}\label{eq:d<c<n<D}
d \le 2^{\Capa} \le \num=\mm+1 \le D + 1. 
\end{equation}
The right most inequality saturates iff the system is classical. If the system is classical, all the inequalities saturate.  
\end{Thm}

Observe the right most and the left most inequality is purely geometric and information theoretic, respectively, 
and these two aspects of the state space are related by the IA inequality \eqref{eq:main}.
Though  $D$, the degree of freedom of the state space, set limits to the information storing capability, it does so only by limiting the asymmetry $\mm$ of the state space.   
So in view of relation between information and geometry, the Minkowski measure $\mm$ of the state space is essential, rather than the diension $\D$.

\subsection{A sufficient condition for $\num=d$}\label{sec:csds}
Though they coincide in classical and quantum theories,
$\num$ is generally greater than $d$. Here we give a
set of natural conditions sufficient for $\num = d$, using the tuple of perfectly distinguishable states, and the group $G$ of affine transforms that leaves the state space invariant (affine bijections on  $\SA$. The former and the latter is a GPT analogue of orthogonal states and reversible operations \cite{ref:NK, ref:KNI}: 

\medskip

\noindent{\bf [T1]} Any $s\in\SA$ is in a convex hull of a maximal set of perfectly distinguishable states $\{s_x\}_{x=1}^d$.

\smallskip

\noindent{\bf [T2]} To each pair of maximal sets of perfectly distinguishable states $\{s_x\}_{x=1}^d$ and 
$\{s'_x\}_{x=1}^d$, there is $g\in G$ that sends $s_x$ to $s'_x$.

\medskip 

In quantum systems, [T1] and [T2] corresponds to the existence of  eigenvalue
decompositions of density operators and transforms between complete sets of orthonormal systems of vectors, respectively.
When $\dim\SA=D= 3$, by the main result of \cite{ref:NK}, these two are enough to characterize quantum and classical systems
\footnote{[T1] and [T2] lead to transitive symmetry \cite{ref:Davies}, and transitive symmetry and [T1] characterizes quantum and classical systems if $D=3$ \cite{ref:NK}.}
. 
Meantime, in higher dimensional systems, a sphere is a clear counterexample to this characterization.

As is shown below, there is a critical state $s_M$ stabilized by affine bijections $G$.
Recall that a state saturating the minimum in \eqref{eq:m} is called a critical state, 
and that the set of critical states, denoted by $\SA^+$, is compact and convex. 
Consider the recursion $(\cdots((\SA^+)^+)^+\cdots)^+$. By \eqref{eq:bound}, this recursion reaches a singleton at finite depth, whose element is denoted by $s_M$.


\begin{Lem}\label{lem:fix}
In any GP model, $\SA^+$ and $s_M$ are stabilized by $G$.
\end{Lem}
\noindent [Proof] 
Suppose $g\in G$. Then by \eqref{eq:ma}, affine invariance of $a_{s^+,s}$, and $g(\SA)=\SA$,
\begin{align*}
&\mm_{g(s^+)}=\max_{s \in \partial \SA} (a_{g(s^+),s})^{-1} = \max_{g^{-1}(s) \in \partial\SA} (a_{s^+,g^{-1}(s)})^{-1} \\
&=\max_{s \in \partial\SA} (a_{s^+,s})^{-1} =\mm_{s^+}.
\end{align*} 
Hence, $s^+\in\SA^+$ implies $\mm_{g(s^+)}=\mm_{s^+}=\mm$, so $g(s^+)\in\SA^+$. 
Analogously, the group $G'$ of affine bijections of $\SA^+$ stabilizes $(\SA^+)^+$. Since $G \subset G'$, $G$ stabilizes $(\SA^+)^+$.
Repetition of this argument shows the second assertion.
\hfill$\blacksquare$

In case of quantum theory, $s_M=\I/d$, so $s_M$ may be viewed as a GPT analogue of the maximally mixed state.

\begin{Prop}\label{prop:fix-cl}
Under [T1-2], for any maximal set of perfectly distinguishable states $\{s_x\}_{x=1}^d$, $s_M$ equals their center
$d^{-1}\sum_{x=1}^d s_x.$ 
\end{Prop}

\noindent [Proof] 
By [T1], there is a maximal set $\{s_x\}_{x=1}^d$ of perfectly distinguishable states whose convex hull, denoted by $K$,  contains $s_M$.
By [T2], for any permutation $\sigma$ on the set $\{1,\cdots,d\}$, there is $g\in G$ with 
$s_{\sigma (x)}=g(s_x)$. Recall the unique common fixed point in the simplex $K$ by all such $\sigma$'s is the center of $K$.
Every $\sigma$, being an element of $G$, fixes $s_M$ by Lemma~\ref{lem:fix}. Therefore, $s_M$ equals the center.

Let $\{s'_x\}_{x=1}^d$ be an arbitrary maximal set of perfectly distinguishable states,
 and suppose $g\in G$ satsfies $g(s_x)=s'_x$. Then 
\begin{align*}
\frac{1}{d}\sum_x s'_x= \frac{1}{d}\sum_x g(s_x)=g(s_M)=s_M.
\end{align*}
The last identity is by Lemma~\ref{lem:fix}. So $s_M$ is the center of $\{s'_x\}_{x=1}^d$ as well.
\hfill$\blacksquare$ 

\begin{Thm}\label{thm:num=d}
Under [T1-2], $\num=d$ and $\SA^+ = \{s_M\}$.
\end{Thm}

\noindent [Proof] 
Let $s$ be an arbitrary state, and suppose it is in the convex hull $K$ of a maximal set $\{s_x\}_{x=1}^d$ of distinguishable states. 
Then 
\begin{align*}
s=\sum_{x=1}^d p_x s_x \le \sum_{x=1}^d s_x=d s_M,
\end{align*}
where the last equality is by Proposition~\ref{prop:fix-cl}. 
Therefore, by \eqref{N-ratio}, $\num\leq d$, which, combined with \eqref{eq:d<c<n<D}, leads to $\num=d$.

Observe any facet of $K$ is perfectly distinguishable from a vertex of $K$, say $s_1$, {i.e.}, there is 
an effect $e$ with $\langle s_1, e\rangle=1$ and $\langle s, e\rangle=0$ for all $s$ on the facet. 
This means any $s$ on this facet is a minimizer of the linear functional $\langle\cdot,e\rangle$ on $\SA$,
and so is on $\partial\SA$. 

Therefore, if $s_b\in\SA^+$ is in $K$,
\begin{align*}
d-1&=\mm(K)\le \mm_{s_b}(K)=\max_{s\in \text{facets of $K$}} (a_{s_b,s})^{-1}\nonumber\\
&\le \max_{s\in \partial\SA} (a_{s_b,s})^{-1}=\mm_{s_b}=\mm=d-1.
\end{align*} 
Therefore, $\mm_{s_b}(K)=d-1$ and $s_b$ is the center of the simplex $K$. 
So by Proposition~\ref{prop:fix-cl} $s_b=s_M$. 
\hfill $\blacksquare$

The latter half of the proof of Theorem~\ref{thm:num=d} essentially proving the following: Suppose $\num=d$. 
Facets of the convex hull $K$ of a maximal set of perfectly distinguishable states are on the boundary of the state space $\SA$.    
Also,  $K\cap\SA_+$ is either empty or a singleton composed of the center of $K$.

Proposition~\ref{prop:fix-cl} and Theorem~\ref{thm:num=d} demonstrates that the postulates  [T1-2] leads to some important properties of quantum and classical systems.
This may have some indication in the characterization of quantum theory.

\section{Conclusion and Discussions}\label{sec:CD}

We have shown the `information induced asymmetry' of state spaces, or the upper bounds of information measures by the asymmetry, 
\eqref{eq:main}, \eqref{eq:d<c<n-F} and \eqref{eq:st-converse}.
The most important relation is the IA-identity \eqref{eq:main}, that linked the geometric characteristic $\mm$ with the information theoretic quantity $\num$:
Others are consequences of this identity and purely information theoretic inequalities.
The dimension of the state space affects information theoretic quantities, but only through the purely geometric upper bound \eqref{eq:bound} of the asymmetry.
So for all GP systems, 
the point-asymmetry is, as far as we know of, the most important geometric factor related to information storing capability. 

Coming back to quantum theory, the point-symmetry of Bloch ball 
and the asymmetry of $d$-level state spaces are `explained' by their information storing capabilities. 

In proving these results, as is often the case, we have assumed 
[R2] in Sec.~\ref{sec:GPT}, feasibility of all the mathematically valid states and measurements. 
But even without this assumption, the weaker version of the statement \eqref{eq:w-IA} and
the information theoretic arguments in Subsec.\ref{subsec:info-measures} are valid,
and they are enough to maintain the above the above argument.
 
A restriction of state to a proper subset $F$ can be considered as a GP system with the state space $F$,
so above consideration indicates a stronger version of the information induced asymmetry: 
{\it Any set $F$ of states capable of storing more than a single bit cannot be point-symmetric,
nor be a subset of any point-symmetric set of states.}

Such an argument shed a light on quantum information theory: 
By \eqref{eq:st-converse-2}, if the image of a quatum channel is point-symmetric, 
its Holevo capacity cannot be greater than a single bit. 
Also, if a set of quantum states contains more than 2 orthogonal states, the set cannot be point-symmetric.

Though we had not mentioned, the Banach-Mazur distance \cite{ref:MS} from the ball is another affine invariant that coincide
with the distance ratio in our preliminary consideration on the quantum and classical systems.
Though this quantity has little information theoretic meaning, 
its coincidence with the Minkowski measure in both classical and quantum systems is noteworthy,
and we conceive this may be related to the manner how the classical system is embedded in the quantum system.
In fact, the conditions [T1-2] in Sec~\ref{sec:csds} concerns a ``classical'' subset of a GP state space,
and the results are more or less related to geometrical properties of a ``classical'' subset.
Our hope is this might be developed to a characterization of quantum systems.


We laid out a GPT version of Shannon theory in 
Sec.~\ref{subsec:info-measures} and Appendices~\ref{sec:pf-d<c<n}-\ref{sec:sig-dim},
but this part is still immature. Our results are mostly about the converse part, and the study of the direct part is left for the future study. 
Also, behavior of information measures in compound systems is another important open problem 
\footnote{The latter will be discussed in our paper in preparation~\cite{ref:MK}.}.
 
{\bf Acknowledgment} 
We would like to thank Profs. M.
Mosonyi, A. Jencova and T. Heinosaari for useful comments and discussions for the early draft of this paper
. This work was supported by JSPS KAKENHI Grant Number JP17K18107 and JP16H01705.

\appendix

\section{General Probabilistic Theories}\label{sec:rep}

A general probability theory contains the triplet of the set of states $\SA$ and measurements $\M$, and the probability rule 
$\Pr[x|M,s]$. Below, we construct its representation on a pair of ordered Banach spaces $\B_*$ and $\B=(\B_*)^*$.
Its motivation is handling of probabilistic mixtures of  states and measurements,
but we do {\it not} presupposes existence of them. Our sole assumption is that  $\Pr[\cdot|M,s]$ is a probability measure.
Instead of taking resource to `convex structure' \cite{ref:Gudder1}, function spaces generated by $\Pr[x|M,s]$'s are used. 

We start from the simple case where measurements takes finitely many values, and then proceeds to the general case.

\noindent {\bf Linear spaces}:   
Represent $s\in \SA$ by the function $\Pr[\cdot|\cdot,s]$, 
and $M\in\M$ by a tuple $(e_x^M)_{x}l$ of functions on $\SA$, where $e^M_x:=\Pr[x|M,\cdot]$, respectively.
This representation naturally induces the equivalence $\sim$
\begin{align*}
s_{1} & \sim s_{2}\Leftrightarrow\Pr[\cdot|\cdot,s_{1}]=\Pr[\cdot|\cdot,s_{2}]\\
M_{1} & \sim M_{2}\Leftrightarrow\forall x\,
\,e^1_x=\Pr[x|M_{1},\cdot]
=\Pr[x|M_{2},\cdot]=e^2_x,
\end{align*}
so the representation is one-to-one up to $\sim$. 
Below and in the main text, we write $s$ for $\Pr[\cdot|\cdot,s]$, by abusing the notation.  Also $\SA$ stands either for a set of vectors or the corresponding quotient set of states modulo $\sim$.  
The notations $M$, $\M$ should be understood similarly. The set of all the effects are denoted by $\E$.
 
Linear combinations of $s$'s and $e^M_x$'s are naturally defined, since they are real valued functions. 
Define the function  $u(s):=1 (s\in\SA)$. Then \eqref{measurement} follows by $u=\sum_x \Pr[x|M,\cdot]$. Also, the identity shows that $u\in\Span\E$. 
Define the paring $\langle\cdot,\cdot\rangle$ 
by the bilinear extension of  the Born-like rule \eqref{eqn:prob} to $\mathrm{span} \SA \times \Span \E$. So the linearity \eqref{eq:mixture} is immediate from the definition
\footnote{If $\SA$ is given as a subset of certain vector space in the beginning, then its original addition and scalar multiplication can differ from those defined above. 
Therefore with respect to this original addition and scalar multiplication, \eqref{eq:mixture} may not hold. But this is not a contradiction.}.

\noindent  {\bf Norms and closure}:
Next, we define the Banach spaces $\B \,(\supset \E)$, $\B^* \supset\B_* (\supset \SA)$. To this end, extend the domain of $e\in\E$ to $\aff \SA$ affinely.  
Let $\T$ be a subset of $\aff \SA$ such that $\T$ contains $\co\SA$ (the convex hull of $\SA$\footnote{The convex hull of the set $A$ is the smallest convex set containing $A$.}), and that any element of $\E$ is positive on $\T$. 

A real-valued function of $\T$ is normed by:
\begin{align}
\norm{f}&:=\sup_{v\in \T}|f(v)|, \label{eq:b-norm}
\end{align}
Let $\LA_{\infty}(\T,\R)$ be the space of functions of $\T$ with $\norm{f}<\infty$, and define $\B$ as its closed subspace generated by $\E$
(here, the domain of each $e\in E$ is restricted to $\T$) for the weak topology given by $\T$
\footnote{This is equivalent to: For any $f\in\B$, finite set $\{v_i\}\subset\T$ and $\epsilon>0$, there is a $f'$ in the linear span of $\E$ with 
$|\langle v_i, f-f'\rangle|<\epsilon$ for all $i$.}.
$\langle\cdot,\cdot\rangle$ is continuously extended to $\Span\T\times\B$, so that the relation \eqref{eq:mixture} extends to $\B$.

$\B$ is a Banach space, since it is norm-closed subspace of  $\LA_{\infty}(\T,\R)$, which is a Banach space 
\footnote{Clearly, the norm closure of $\B$ is the subset of the weak closure of it. The latter is $\B$ itself by definition, so is the norm closure of $\B$. Also, the positive cone $\B_+$ is norm closed by the definition \eqref{eq:b-ord}.}. 

Identify $v\in\T$ with a linear functional $\langle v,\cdot\rangle\in\B^{\ast}$, where $\B^{\ast}$ is the dual of $\B$ (with the norm topology) and normed by \eqref{eq:normE}.
Define $\B_\ast$ as the norm-closure of the linear space generated by this natural embedding of $\T$ into $\B^{\ast}$. Extend  $\langle \cdot, \cdot\rangle$ continuously to $\B\times\B_*$, , so that the relation \eqref{eq:mixture} is valid on this extended domain.

\noindent  {\bf Duality}: 
To show that $(\B_*)^*=\B$ holds, by the main result of \cite{Kaijser},  we only have to show that {\it (i)} $\T$ separates elements of $\B$ and {\it (ii)} the closed unit ball $U\in\B$ is compact for the weak topology given by $\T$. {\it (i)} is clear by definition, and {\it (ii)} is by Tychonoff's theorem
\footnote{The weak topology given by $\T$ can be viewed as the product topology of ${\R}^{\T}$. 
By Tychonoff's theorem, the set $[-1,1]^{\T}$ is compact for this topology, so its intersection (call it $\tilde{U}$) with the weakly closed set $\B$ (here, the domain of each $e\in E$ is restricted to $\T$) is compact. Clearly, $\tilde{U}$ and $U\subset\B$ are homeomorphic, so $U$ is compact.}. 

Since $(\B_*)^*=\B$, the representation \eqref{eq:sup-norm-2} of the norm $\norm{\cdot}_1$ is valid
\footnote{See Corollary III.6.7 of \cite{ref:Conway}}.

If $A$ is a convex subset of $B_*$, the closure by the norm and by the weak topology given by $\B$ ($\sigma(\B_*,\B)$-topology) are identical \footnote{See Theorem V.1.4. of \cite{ref:Conway}.}. 
So we simply denote it by $\overline{A}$. 

\noindent {\bf Order structures}:
The positive cone $\B_{+}$ (the set of positive elements in $\B$) is defined by
\begin{align}
f\ge 0&\Leftrightarrow \forall v\in\T\, f(v)\ge 0. \label{eq:b-ord},
\end{align}
and the positive cone $\B_{*+}$ is the closure of the cone generated by $\T$. 

One of the duality relations \eqref{eq:order-dual-2} is clear by definitions, and the other \eqref{eq:order-dual-1} is by the Hahn-Banach theorem
\footnote{See e.g.  23.1 of \cite{ref:KellyNamioka}. For the reader's convenience, we write the proof here. 
$\Rightarrow$ is clear. To show $\Leftarrow$, suppose  $v_0\not\in \B_*$. 
Then by the Han-Banach theorem, there is $f\in\B$ and $c$ with $\langle v_0,f\rangle<c\le\langle v,f\rangle$ for all $v\in\B_{*,+}$.
If $v_1\in\B_{*,+}$, then $\lambda v_1\in\B_{*,+}$ for all $\lambda>0$. So $\langle v_1,f\rangle\ge 0$. Since $0\in\B_{*+}$,  one may choose So let $c=0$.
Therefore,  $\langle v_0,f\rangle<0$. This implies $\Leftarrow$.}.   

Clearly,  to each given element $f$, there is an $n\in\N$ with $-nu\le f\le nu$ ($u$ is an {\it order unit} of $\B$). 
$u$ is an interior point of $\B_+$, since for any $f\ge 0$ with $\|f\|<1/2$, $u+f\ge 0$. 
Meantime, $\B_{*+}$ may have no interior point. For example, the space of infinite dimensional quantum density operators has none.  

$\B_+$ is pointed ($\B_+\cap(-\B_+)=\{0\}$), and so is $B_{*+}$.
\footnote{Suppose $f\\B_+\cap(-\B_+)$. Then $f(v)=\langle v,f\rangle= 0$ holds for any $v\in\T$. So $f=0$ by definition. The proof for $\B_{*+}$ is almost parallel.}. 

Moreover, $\B_+$ is generating ($\B_+-\B_+=\B$), since $f$ equals the sum $\frac{1}{2}(f+\|f\|u)+\frac{1}{2}(f-\|f\|u)$, where  $(f+\|f\|u)\in\B_+$ and $(f-\|f\|u)\in -\B_+$. 
$\B_{*+}$ is generating also, as is proved later.

\noindent {\bf Relations between norms and orders}:
Since $\langle v,u\rangle=1$ for all $v\in\T$, by  \eqref{eq:b-norm} and \eqref{eq:b-ord},  
the norm $\norm{\cdot}$ satisfies  \eqref{eq:ordnorm}
\footnote{The norm satisfying \eqref{eq:ordnorm} is called the order unit norm with respect to $u$See pp. 8, 23-26 of \cite{ref:OrderedVec}.}
 
Observe the closure $\overline{\T}$ of $\T$ is identical to 
\footnote{To show the identity of two sets, observe that $\|e\|=1$ implies $e\le u$ by \eqref{eq:b-ord}. Therefore, if $v\in\B_*$ is positive,  
$\langle v, e\rangle$ cannot be larger than $\langle v, u\rangle$, so $\|v\|_1=\langle v, u\rangle$, and the two sets are identical.}   
\begin{align}
\overline{\T}&= \{v\in\B_{*,+}; \langle v, u\rangle=1\}\nonumber\\
                 &=\{v\in\B_{*,+}; \|v\|_1=1\} \label{eq:T-dense}
\end{align}
Therefore, rewriting the range of the supremum in \eqref{eq:b-norm}, we obtain two identities in \eqref{eq:sup-norm}.
 
The closed unit ball $\mathrm{ball}\B_*$ in $\B_*$ is given by 
\footnote{ 
This is a special case of Theorem 3.2, p.18 of \cite{ref:OrderedVec}, but here we give an elementary proof.
`$\supset$' is clear since  $\mathrm{ball}\B_*$ is a closed convex set containing  $\overline{\T}$ and $-\overline{\T}$. 
To see `$\subset$', suppose $v_0\not\in A$, where $A$ is the RHS of \eqref{eq:ball}.
Since $A$ is convex and closed, by the Hahn-Banach theorem, 
there is $f$ with $\langle v_0,f\rangle>c$ and $\langle v,f\rangle\le c$ for all $v\in A$.
Since by \eqref{eq:T-dense} and \eqref{eq:sup-norm},  $\norm{f}=\sup_{v\in A}\langle v,f\rangle\le c< \langle v_0,f\rangle\le \norm{f}\norm{v_0}_1$. 
Therefore,  $\norm{v_0}_1>1$. Therefore, $A$ should be identical with $\mathrm{ball}\B_*$.}
\begin{align}
\mathrm{ball}\B_*=\overline{\co(\overline{\T}\cup(-\overline{\T})}.\label{eq:ball}
\end{align}

By the lemma in p.18 of \cite{ref:OrderedVec}, $\co(\overline{\T}\cup(-\overline{\T})$ contains the interior of $\mathrm{ball}\B_*$, 
which in turn contains  $\frac{1}{1+\epsilon}\mathrm{ball}\B_*(\forall\epsilon>0)$. 
Therefore, to each given $v\in\B_*$ and $\epsilon>0$, we can find the decomposition $v_+$ and $v_-$ such that 
\begin{align}
&v=v_{+} - v_{-}, v_{+}, v_{-}\in\B_{*+},  \nonumber\\
&\norm{v_{+}}_1+\norm{v_{-}}_1=(1+\epsilon)\norm{v}_1. 
\end{align} 
So the cone $\B_{*+}$ is generating, $\B_{*}=\B_{*+} -\B_{*+}$.
If $\epsilon=0$,  $v_{+} - v_{-}$ can only approximate $v$ up to arbitrary precision.
$\epsilon=0$ and the strict identity is possible if  $v_+$ and $v_{-}$ are positive elements of $\B^*$, rather than of $\B_*$
\footnote{Observe $\norm{v}_1$ is given by the SDP $\sup\{\langle v,f\rangle; -u\le f\le u\}$. 
Since the set $\{f; -u\le f\le u\}$ is a closed convex set having an interior point and the optimal value is finite,
by Theorem 8.6.1 of \cite{ref:Luenberger},  the dual and the primal problem has equal optimal value, and the infimum in the dual problem is achieved. 
The dual problem is $\min_{v_+,v_{-}\ge 0}\sup_{e}\{\langle v-(v_+-v_{-}),e\rangle+\langle v_+ + v_{-}, u\rangle\}$, which equals 
 $\min\{\langle v_+ + v_{-}, u\rangle; v= v_+-v_{-}, v_+,v_{-}\ge 0\}$. A solution to this program $(v_+,v_-)$ is a desired decomposition.
}
\footnote{But the decomposition $v=v_+-v_{-}$ is not unique, even if $\dim\B<\infty$. 
An example is the space of binary classical channels. This space can be viewed as a GPT state space, 
while each  effect is a pair of an input probability distribution and a positive function on output bits with unit sup-norm.
An $s\in\SA$ can be represented by the quadruple of real numbers $(b,1-b, c,1-c)$, 
so an element of $\B_*=\B^*$ is represented by a quadruple $(a,b, c,a+b-c)$, 
and its norm is  $\max\{|a|+|b|,|c|+|a+b-c|\}$. It is positive iff all is  components are positive.
Then  $(-2,2,-1,1)$ admits a family of decompositions $(0,2,1-t,1+t)-(2,0,2-t,t)$, where $t\in[0,1]$.}    

\noindent {\bf Measurements over measurable spaces}:
To generalize the above construction to measurements over a measurable space $(\Omega,\sigma(\Omega))$, replace $e_x$ by $e[B]:=\Pr[B|M,\cdot]$, 
where $B\in\sigma (\Omega)$. Then $u=\Pr[\Omega|M,\cdot]$. Then all the previous arguments clearly go through.

 It remains to show $\sigma$-additivity of $e[\cdot]$ in the weak sense:  For any $v\in\B_*$. 
 \begin{align}
 \langle v, e[\cup_{i=1}^{\infty}B_i]\rangle=\sum_{i=1}^\infty \langle v, e[B_i]\rangle
 \end{align}
Clearly,  it is true if  $v\in\SA$, and the statement is easily extended to $v\in\mathrm{span}\SA$. So we only have to extend it to $v\in\B_*$. Clearly, $e[\cdot]$ is finitely additive. So, 
\begin{align*}
&|\langle v, e[\cup_{i=1}^{\infty}B_i]\rangle -\sum_{i=1}^n \langle v, e[B_i]\rangle|= |\langle v, e[\cup_{i=n}^{\infty}B_i]\rangle |\nonumber\\
&\le |\langle v', e[\cup_{i=n}^{\infty}B_i]\rangle |+\norm{v-v'}_1 \norm{e[\cup_{i=n}^{\infty}B_i]}\nonumber\\
 &= |\sum_{i=n}^\infty \langle v', e[B_i]\rangle|+\norm{v-v'}_1\norm{e[\cup_{i=n}^{\infty}B_i]}. \nonumber 
\end{align*} 
Here the last `$=$' is by the $\sigma$-additivity of $\langle v', e[\cdot]\rangle$, $v'\in\mathrm{span}\T=\mathrm{span}\SA$. 

So if the first term of the last end vanishes as $n\to\infty$, the proof completes by 
taking the limit first by $n\to\infty$ and then by $\norm{v-v'}_1\to 0$.
To show this, observe $\sum_{i=1}^\infty \langle v', e[B_i]\rangle$ is finite, being equal to $\langle v', e[\cup_{i=1}^\infty B_i]\rangle$.
So  $\sum_{i=n}^\infty \langle v', e[B_i]\rangle$ vanishes as $n\to\infty$, completing the proof. 


\noindent {\bf Integral}:
$\langle v, e[\cdot]\rangle$ is a signed measure for any $v\in\B_*$, and the integral $\int g(x) \langle v, e[dx]\rangle$ is well-defined if $g$ is a measurable function.
If  this integral is finite for all $v$ in $\B_*$, the linear map $v\to\int g(x) \langle v, e[dx]\rangle$ is bounded by the principle of uniform boundedness \footnote{e.g., Theorem 14.1 of \cite{ref:Conway}}. Therefore,  an element $f\in\B$ satisfies $\int g(x) \langle v, e[dx]\rangle=\langle v, f\rangle$ for all $v\in\B_*$. $f$ may be denoted by $\int g(x) e[dx]$.

\noindent {\bf Arbitrariness of order structures}: 
The order structures that are introduced by our recipe may not be unique, due too the arbitrariness of $\T$.
For example, suppose  a $v_0\in\aff\SA$  is strictly separated from $\SA$ by an $e_0\in\E$ and that any $e\in\E$ is positive on $v_0$.
Then  $\T=\co(v_0\cup\SA)$ and $\T=\co\SA)$ will introduce different norm and order structures. 
If such $v_0$ exists, an $f\in\Span\E$ is strictly separated from the cone generated by $\E$ by some element of $\Span\T$  and is positive on $\T$.
For example, if some  pure states or POVMs in quantum theory are prohibited (by some fundamental physical law), this situation occurs.

Such arbitrariness does not exist if $\SA$ and $\E$ are so large that the cone generated by the latter is dense in the dual cone of the former.
This is the case if (but not only if) all the affine functionals  on $\co\SA$ taking values in $[0,1]$ are in the set $\E$.

\noindent {\bf Physical postulates, extension of state and effect space}:  
In construction of the vector representation, our only assumption is that each $\Pr[\cdot|s,M]$ is a probability measure.
But in its physical interpretation, some more assumptions are necessary. Below, we sketch some of them.

In our main text,  GP systems are treated as black boxes in the classical system. 
The output of the measurements are classical signals, and can be post-processed classically. 
Several GP systems can be prepared in parallel, labeled, and arranged. 

Also, {\it probabilistic mixtures of states and measurements are physically feasible,  
and the data of resulting from the mixtures obeys probability distributions given by RHSs of \eqref{eq:mixture}.}

If this postulate is accepted, it is convenient (but not necessary) to include convex mixtures of states and effects in $\SA$ and $\E$, respectively.
Following suit with majority of authors, we accepted these assumptions in the main text.

Taking the closure of  $\SA$ (by $\sigma(\B_*, \B)$-topology) and $\E$ (by $\sigma(\B, \B_*)$-topology) is mathematically convenient and physically harmless, 
since the difference between $\SA$ and its closure is almost impossible to detect    
\footnote{Extension of $\SA$ to $\B^*$ can be problematic, regardless its mathematical convenience.
Recall its equivalence in classical theory is not a probability measure in usual sense ($\sigma$-additivity of events fails, 
so the `probability' of the joint of the mutually exclusive events might not equal the sum of `probability' of the each event.). Similar problems occur in its quantum equivalence.}.

In this paper, we did not give distinct symbols to physically indistinguishable states, so `a state $s$' in fact refers to an equivalence class.
The notion `a measurement $M$' should be understood analogously. In other words, it is almost postulated that
{\it the physically indistinguishable states/measurements are identified}.

\noindent {\bf Common technical assumptions}: 
Usually, it is supposed that the cone generated by $\SA$ and $\E$ is identical with $\B_{*+}$ and $\B_{+}$, respectively.
The stronger assumption ([R2] in our paper), which states any mathematically valid states and measurements are physically feasible, are frequently postulated as well.
But there is no compelling arguments for them other than their common use and fruitfulness of their consequences. 
It is also common to assume that $\dim\B<\infty$ ([R1] in our paper) for technical simplicity.

\section{An expression of $\mm$}\label{sec:mm-another}
 (3.1.2) of \cite{ref:MS} reads: 
\begin{align}
(\mm_{s_0}+1)^{-1}
= \min_{e \ge 0 , \norm{e}=1} \langle s_0, e\rangle. \label{eq:t-e}
\end{align}
In a quantum system, this is the minimization of the overlap between the given state $s_0$ and a pure state. 

For completeness, the proof is given here.
By \eqref{eq:mm-gp}, 
\begin{align*}
&(\mm_{s_0}+1)^{-1} = \max\{ c^{-1} \ ; \  s_0\ge c^{-1} s \}\\ 
      &= \max\{ c^{-1} \ ; \ \forall e\in\B_+\, \norm{e}=1, \langle s_0, e\rangle \ge c^{-1} \langle s,e\rangle \} \\
     &= \min_{e\ge 0, \norm{e}=1} \max\{ c^{-1} \ ; \ \langle s_0, e\rangle \ge c^{-1} \langle s,e\rangle \} \\
     &=\min_{e\ge 0, \norm{e}=1} \langle s_0, e\rangle (\langle s, e\rangle)^{-1}.
\end{align*}
Therefore , 
\begin{align*}
(\mm_{s_0}+1)^{-1}
= \min_{e \ge 0 , \norm{e}=1} \langle s_0, e\rangle (\max_{s\in \partial \SA}\langle s, e\rangle)^{-1} 
\end{align*}
This equals \eqref{eq:t-e} by $\norm{e}=\sup_{s\in \partial \SA}\langle s,e\rangle$, which follows from
\eqref{eq:sup-norm}, \eqref{eq:S} and $e\ge 0$. 

In case of quantum systems, \eqref{eq:t-e} is extremely useful. Indeed, it is immediate that 
\begin{align*}
(\mm_{s_0}+1)^{-1} =\lambda_{\min}, \ \ \mm_\rho = (\lambda_{\min})^{-1} - 1
\end{align*}
where $\lambda_{\min}$ is the smallest positive eigenvalue of $\rho_0$.
 (The first identity also appears in \cite{ref:J}.)

\section{The dual representation of $\num(F)$}\label{sec:proof}

In this subsection, we deal with the slightly generalized version of \eqref{N-ratio}, where  $F$ is not limited to $\SA$ itself. 
This is equivalent to $\num(\SA)$ under the assumption that $\SA$ may be a proper subset of the RHS of \eqref{eq:S} but all the measurements with \eqref{measurement} are feasible.

Theorem 2.1 of \cite{ref:Detection}, though stated in a quite generalized form, reads in our notation and context,
\begin{align}
\max_{M} \sum_{x=1}^l \langle s(x),e_x \rangle= \min_{s_0\in\SA}\min\{c ; \forall x \ cs_0 \ge s(x)\}. \label{eq:KRF}
\end{align}
Also similar identities appear in various references such as \cite{ref:Detection-2, ref:KRF}.
The LHS appears close to \eqref{primal}, but $s(\cdot)$ and $l$ are fixed.
One can prove \eqref{N-ratio} from \eqref{eq:KRF}, e.g., by approximating the continuous set by a finite subset.
But here we show that slight modification of the proof of \eqref{eq:KRF} leads to \eqref{N-ratio}.

First, we review the proof of \eqref{eq:KRF}. Relax the constraint on $(e_x)_{x=1}^l$ to 
\begin{align}
e_x\ge 0, \ \sum_{x=1}^l e_x\le u \label{eq:sub-measurement}.
\end{align}
This relaxation does not change the maximum, since given $(e_x)_{x=1}^l$ with \eqref{eq:sub-measurement},
a measurement $(e'_x)_{x=1}^l$ defined in the sequel is at least as good as $(e_x)_{x=1}^l$:
Let $e'_1:=e_1+(u-\sum_{x=1}^l e_x)$ and $e'_x:=e_x \ (x\neq 1)$. Then 
$\sum_{x=1}^l \langle s(x),e_x \rangle \le \sum_{x=1}^l \langle s(x),e'_x \rangle$,
as $e'_x\ge e_x$ for all $x$.

After this rewriting, which is necessary to satisfy the condition (ii) below, 
the strong duality of SDP \cite{ref:Detection, ref:Detection-2, ref:Luenberger} 
is applied for the maximization of $(e_x)_{x=1}^l$, leading to \eqref{eq:KRF}.

In the proof of \eqref{N-ratio}, we rewrite the problem further before application of the duality. 
Let $\NF$ be the set of all the tuples $N=(e_s)_{s\in F}$ 
of elements of $\B_+$ such that $e_s=0$ except perhaps finitely many points. 
Then $s(\cdot)$ is expressed by the support of $N=(e_s)_{s\in F}$: 
If $s(x)$'s are elements of $\{s;e_s\neq 0\}$ and $l=|\{s;e_s\neq 0\}|$,
\begin{align*}
l\cdot P_{suc}(s(\cdot),N)= \sum_{s\in F} \langle s,e_s\rangle=: g(N),
\end{align*}
which is well-defined on $\NF$. Therefore, 
\begin{align}\label{eq:NF}
\num(F)=\sup_{N\in \NF,G(N)\ge 0}g(N),
\end{align}
where $G(N):=u- \sum_{s\in F} e_s $.
This is a linear optimization problem with semidefinite constraint. 
So by the strong duality, the proof completes.

Let us check the sufficient condition for the strong duality (Theorem 8.6.1 of \cite{ref:Luenberger}, Theorem 2 of Appendix II of \cite{ref:Detection}).

\smallskip

 \noindent (i) $-g$ and $-G$ are convex functions, $\NF$ is a convex set\footnote{$\NF$ need not to be topologized.}.

 \noindent (ii) The image of $G$ is an ordered normed vector space, and there is $N_1\in \NF$ such that $G(N_1)>0$. 
 
 
 \noindent (iii) $\num(F)$ is finite.

\smallskip


(i) is clear by their definitions: 
Recall that the cardinality of the support of $(e_s)_s$ is arbitrary, as long as it is finite.

The former half of (ii) is clear, since $G$ is a map into $\B$. 
To see its latter half, define $N_1 \in \NF$ by $e_{s_0}^1=u/2,$ and $e_s^1=0$ ($s\neq s_0$). 
Then $G(N_1)=u/2$ is in interior of $\B_+$ 
\footnote{As $\B_{*+}$ is generated by $\{v\in\B_{*+} ; \langle v,u\rangle=1\}$, $\langle v,u/2\rangle>0$ for any non-zero $v\in\B_{*+}$. 
So this follows by the separation theorem \cite{ref:ConvAna}. Recall $\B_{*}=\B^*$ in our case. 
If $\dim \B=\infty$, use the representation of $\B$ by functions over $\SA$ in Appendix~\ref{sec:rep}. 
Then $u/2$, the function which is constatanly 1/2, is clearly in the interior point of the positive cone.}.
That $\NF$ is not topologized is immaterial here, as its image by $G$ is of our concern. 
 
To see (iii), it suffices to show the RHS of \eqref{N-ratio} is finite by the weak duality \eqref{eq:weak-dual}. 
Clearly, it suffices to show the statement for $F=\SA$. 
In the proof of Theorem~\ref{thm:main}, we had shown this quantity equals $\mm+1<\infty$, so (iii) is confirmed
\footnote{In the infinite dimensional case, $\num=\eqref{eq:dual}$ is valid if the range of $\num$ is allowed to take $\infty$: If $\num<\infty$, the strong duality implies the identity. If $\num=\infty$, the weak duality implies $\eqref{eq:dual}=\infty$.}.
((iii) is trivial in case of \eqref{eq:KRF}.)

For completeness, we compute the dual problem: 
\begin{align}
& \min_{\xi\in\B_+^\ast}\sup_{N\in \NF}\{ g(N)+\langle \xi,G(N)\rangle \} \nonumber\\
 &=\min_{\xi\in\B_+^\ast} 
 \{ \langle \xi,u\rangle+\sup_{N\in\NF}\sum_{s\in F}\langle s-\xi,e_s\rangle \}\nonumber\\
&= \min\{ \langle \xi,u\rangle\ ; \ \xi\in\B_+^\ast, \ \forall s\in F,\ s\le \xi \rangle \}. \label{eq:dual}
\end{align}
Here, the last `=' is shown as follows. Since $s \not\le \xi $ implies $\sup_{e_s\ge 0}\langle s-\xi,e_s\rangle=\infty$, $\xi$ should satisfy $s\le\xi (\forall s\in F)$ to achieve the infimum. 
Under this constraint, for example, $e_s$=0 achieves the supremum about $e_s$
\footnote{
In fact, there is $(e_s)_{s\in F}$ with 
$\langle s-\xi, e_s \rangle =0$ and $\sum_{s\in F} e_s=u$, and such 
$(e_s)_{s\in F}$ gives an optimal measurement.
}.

\eqref{eq:dual} with $F=\SA$ is identical to \eqref{N-ratio}: By $\B_*=\B^*$,
there is a normalized ($\langle s_0,u\rangle=1$) positive element $s_0$ of $\B_{*+}$ such that $\xi = c s_0$,
so $\left\langle \xi ,u\right\rangle = \left\langle c s_0, u\right\rangle = c$. 

\section{Max-relative entropy} \label{sec:pf-d<c<n} 
Below we present a proof of \eqref{eq:d<c<n-F}, which is shorter but uses the assumption [R2]. 
Recall in the proof in the main text, this assumption was not used.

Here, we combine \eqref{N-ent} and known relations between $\Di_{\max}$ and relative entropy:
\begin{align}
&\Di(p\Vert q)\leq \Di_{\max} (p\Vert q), \label{eq:Dm>D} \\
&\Di_{\max} (p^M_{s_1}\Vert p^M_{s_2})\leq \Di_{\max} (s_1\Vert s_2) \label{eq:Dm>Dm}
\end{align}
where $p$, $q$ are probability distributions \cite{ref:Dmax} 
\footnote{Though in \cite{ref:Dmax} they discusses quantum systems only, the generalization to GPTs is trivial.}.

These lead to the second inequality of \eqref{eq:d<c<n-F}: 
\begin{align*}
&\inf_{p_0} \sup_{x} \Di(p^M_{s(x)}\Vert p_0)
 \underset{\mathrm{(i)}}{\leq} \inf_{s_2\in\SA} \sup_{s_1\in \SA} \Di(p^M_{s_1} \Vert p^M_{s_2}) \\
&\underset{\mathrm{(ii)}}{\leq} \inf_{s_2\in \SA} \sup_{s_1\in \SA} \Di_{\max}(p^M_{s_1} \Vert p^M_{s_2})
\underset{\mathrm{(iii)}}{\leq} \inf_{s_2\in \SA} \sup_{s_1\in \SA} \Di_{\max} (s_1\Vert s_2)
\end{align*}
where (i) is by the comparison of the ranges of the variables, (ii) is by \eqref{eq:Dm>D}, and 
(iii) is by \eqref{eq:Dm>Dm}. Therefore, taking supremum about $s(\cdot)$ and $M$, we obtain the desired inequality.



\section{Signaling dimension}\label{sec:sig-dim}
Recently \cite{ref:FrenkelWeiner}, another information measure, {\it signaling dimension}, which we denote by $\sig$, is introduced. It is the smallest integer having the following property:
To each given $(s_x)_{x}$ and $(e_y)_{y}$, there are a probability distribution $R_b$ over $b$'s and transition probabilities $P_{y|z}^b$, $Q_{z|x}^b$ $(z\in\{1,\cdots,\sig\})$ such that 
\begin{align*}
\langle s_x,e_y\rangle=\sum_{b} \sum_{z=1}^{\sig} P_{y|z}^b Q_{z|x}^b R_b
\end{align*}

Different from other information measures, this new one quantifies the number of classical noiseless channel necessary for `simulation' of a given physical system by a classical channel. By this intuition, 
\begin{align}\label{eq:si<num}
\sig\ge\num
\end{align}
is expected. The proof runs as follows. 
\begin{align*}
&\sum_x \langle s_x,e_x \rangle
=\sum_x\sum_{b} \sum_{z=1}^{\sig} P_{x|z}^b Q_{z|x}^b R_b \\
&\leq \max_{b} \sum_x \sum_{z=1}^{\sig} P_{x|z}^b Q_{z|x}^b 
\leq \max_{b} \sum_x \sum_{z=1}^{\sig} P_{x|z}^b \\
&= \max_{b} \sum_{z=1}^{\sig} 1 =\sig,
\end{align*}
where the first inequality hods since $R_b$ is a probability distribution. 
Taking maximum over all $(s_x)_x$ and $(e_x)_x$, the asserted inequality is proved.

\section{Helstrom family}\label{sec:HF}
Here, the conjectured existence of the Helstrom family of ensembles \cite{ref:KMI} is shown.

Below, $F = (s_x)_{x=1}^l$ is a finite family of states, and $p=(p_x)_{x=1}^l$ is a probability distribution on it. 
Consider maximization of the success probability of state detection 
\begin{align}
\num(F,p)&:=\sup_{M}\sum_{x=1}^l p_x \langle s_x, e_x \rangle\nonumber\\
&= \min \{\langle \xi, u \rangle ; \xi \ge p_x s_x, x=1,\cdots,l \}. \label{eq:stDual}
\end{align}
where $M$ runs over all the measurements having values in $\{1,\cdots,l\}$,
and the second identity is by the strong duality of the linear SDP\cite{ref:Detection}.

A {\it weak Helstrom family} \cite{ref:KMI} is a family of ensembles 
$\{\tilde{p}_x, s_x, \tilde{s}_x \}_{x=1}^l \ (\tilde{p}_x \in [0,1], \tilde{s}_x \in \SA)$ such that
$p_x/\tilde{p}_x=:r$ and $\ \tilde{p_x} s_x + (1-\tilde{p}_x) \tilde{s}_x=:s_0$ are constant of $x$. 
This $r$ is called a {\it Helstrom ratio}, and it holds that 
$r\ge \num(F,p) $. 
The family is called a {\it Helstrom family} if $r=\num(F,p)$. 

\begin{Thm}\label{thm:HF}
A Helstrom family exists to every $F$ and $p$.
\end{Thm}
\noindent [Proof] Suppose $\xi\ge p_x s_x$ for all $x$. 
As easily checked, $r:=\langle \xi,u\rangle$, $s_0:=r^{-1}\xi$, $\tilde{p}_x:=r^{-1}p_x$, and $\tilde{s}_x:=(1-\tilde{p}_x)^{-1}(s_0-\tilde{p}_x s_x)$ defines a weak Helstrom family 
($1\ge\tilde{p}_x $ and $\tilde{s}_x\ge 0$ are by $\xi\ge p_x s_x$).
Thus, if $\xi \ge 0$ is the minimizer of \eqref{eq:stDual}, it is a Helstrom family.
\hfill $\blacksquare$

\end{document}